\documentclass[epj]{svjour}
\usepackage{graphics}
\usepackage{amsmath}
\usepackage{natbib} 
\usepackage{graphicx}
\usepackage{subcaption}
\usepackage{tikz}
\usepackage{microtype}
\usepackage{tabularx}
\usepackage{booktabs}
\begin{document}
\title{Enhancing the Angular Resolution of Large Array of imaging atmospheric Cherenkov Telescope (LACT) at Ultra-High Energies}
\subtitle{}

\author{
Zhipeng Zhang\inst{1} \and 
Ruizhi Yang\inst{1,2,3,6}\thanks{\emph{Corresponding author email:} yangrz@ustc.edu.cn} \and 
Shoushan Zhang\inst{4,5,6} \and 
Jiali Liu\inst{4,5,6} \and 
Liqiao Yin\inst{4,5,6} \and 
Yudong Wang\inst{4,5,6} \and 
Lingling Ma\inst{4,5,6} \and 
Zhen Cao\inst{4,5,6}
}                          
\offprints{}          
\institute{
School of Astronomy and Space Science, University of Science and Technology of China, 96 Jinzhai Road, Hefei, 230026, Anhui, China \and
CAS Key Laboratory for Research in Galaxies and Cosmology, Department of Astronomy, University of Science and Technology of China, 96 Jinzhai Road, Hefei, 230026, Anhui, China \and
Deep Space Exploration Laboratory / School of Physical Sciences, University of Science and Technology of China, 96 Jinzhai Road, Hefei, 230026, Anhui, China \and
Key Laboratory of Particle Astrophysics, Institute of High Energy Physics, 19B Yuquan Road, Shijingshan District, Beijing, 100049, China \and
Department of Physics, University of Chinese Academy of Sciences, 19A Yuquan Road, Shijingshan District, Beijing, 100049, China \and
Tianfu Cosmic Ray Research Center, Chengdu, Sichuan, China
}
\date{Received: date / Revised version: date}
%
\abstract{
The Large Array of Imaging Atmospheric Cherenkov Telescopes (LACT) is dedicated to high-resolution morphological studies of PeVatrons. In this work, we present a fundamental investigation into stereoscopic direction reconstruction for the LACT array, specifically addressing the challenges of ultra-high-energy observations. {We demonstrate that the standard Hillas parameterization introduces a significant reconstruction bias under severe image leakage. To mitigate this, we introduce an approach utilizing a 2D Gaussian fit, achieving an exceptional angular resolution of better than $0.06^\circ$ at $100\text{ TeV}$ within the central $0^\circ\text{--}1^\circ$ offset bin, and maintaining better than $0.12^\circ$ across offsets up to $4^{\circ}$.} Building on this robust baseline, we evaluate advanced weighting schemes by utilizing a LightGBM-based quantile regression model to independently estimate single-image quality. Applying these quality-based weights yields a consistent improvement of $0.02^\circ$ to $0.03^\circ$ for high-energy, large-offset events using both the \textit{HillasWeightedSum} and \textit{HillasWeightedDisp} methods. Finally, we explore a pixel-wise likelihood reconstruction technique utilizing Neural Ratio Estimation. While its practical realization depends heavily on minimizing the gap between Monte Carlo simulations and observational data, this exploratory approach demonstrates the potential to yield an overall improvement of approximately 15\% to 35\% at $100~\rm TeV$ across the entire field of view. Such high angular resolution is critical for disentangling complex emission regions and mapping the internal structures of PeVatrons.
\PACS{
      {95.55.Ka}{X- and $\gamma$-ray telescopes and instrumentation} \and
      {95.85.Pw}{$\gamma$-ray astronomy}
     } 
} 
\maketitle
\section{Introduction}
\label{sec1}

Recent discoveries by the Large High Altitude Air Shower Observatory (LHAASO) have fundamentally reshaped our understanding of the ultra-high-energy (UHE) universe. The detection of UHE gamma-ray emission from sources such as the Cygnus Cocoon \cite{lhaaso2024ultrahigh} and SS 433 \cite{cao2025ultrahigh} demonstrates that the Milky Way is populated with PeVatrons capable of accelerating particles to PeV energies. However, while air shower arrays like LHAASO possess superior sensitivity at UHE regimes, their angular resolution is inherently limited. This limitation hinders the detailed morphological study of complex regions where multiple emitters may be confused.

To disentangle these complex regions and probe the precise locations of particle acceleration, Imaging Atmospheric Cherenkov Telescopes (IACTs) with superior angular resolution are essential. While current generation IACTs (e.g., H.E.S.S. \cite{aharonian2006observations}, MAGIC \cite{albert2008vhe}, VERITAS \cite{weekes2002veritas}) provide excellent resolution, they lack the effective area required to effectively detect the low fluxes characteristic of the UHE regime. The Large Array of Imaging Atmospheric Cherenkov Telescopes (LACT) \cite{zhang2025performance}, consisting of 32 telescopes to be co-located with the LHAASO array, is designed to bridge this gap. With an effective area reaching the $\text{km}^2$ level, LACT aims to perform detailed morphological studies of the PeVatrons discovered by LHAASO.

A significant challenge for next-generation IACTs lies in the nature of these UHE sources: the majority are spatially extended, spanning degrees on the sky. Observing extended sources with IACTs is technically demanding. As noted in previous studies \cite{celli2024detection,zhang2024prospects}, the sensitivity degrades with increasing source size due to the larger integration region required, which introduces much more cosmic-ray background. Furthermore, standard background modeling techniques \cite{berge2007background} (e.g., Reflected Region Background and Ring Background) become difficult to apply when the source emission fills a significant portion of the camera's field of view (FoV).

Despite these challenges, resolving internal substructures remains a scientific imperative, particularly for identifying hotspots and conducting spatially resolved spectral analysis of acceleration sites \cite{abdalla2018hess}. For LACT, its co-location with LHAASO provides a unique and decisive advantage in this regard. By leveraging LHAASO's KM2A muon detectors, LACT can achieve nearly background-free observations above $100\text{ TeV}$ at small zenith angles \cite{zhang2024prospects}. This powerful synergy makes LACT exceptionally well-suited for deep, long-term observations of complex, extended sources compared to standalone IACTs. Consequently, the primary limiting factor for LACT's scientific reach in this regime is no longer background suppression, but rather the strict requirement for a uniformly excellent angular resolution across the entire Field of View (FoV) to map these extensive structures without morphological distortion.

However, maintaining this uniform, high-performance angular resolution presents a significant challenge for LACT due to its extreme high-altitude deployment ($\sim 4410\text{ m}$ a.s.l.). At this elevation, the extensive air shower maximum occurs much closer to the telescopes, producing significantly larger Cherenkov images on the camera focal plane compared to lower-altitude instruments. This proximity inevitably leads to severe image leakage (truncation) at the camera edges, which notably deteriorates the image parameterization and subsequent reconstruction performance, particularly for ultra-high-energy events and at large off-axis angles.

While executing observations at Large Zenith Angles (LZA) can partially mitigate this truncation by increasing the atmospheric slant depth -- yielding more compact images and an expanded effective area at the highest energies \cite{zhang2024layout}. This approach fundamentally limits the observable duty cycle. More importantly, standard small zenith angle observations are highly preferred to maximize the critical joint-observation synergy with KM2A. Therefore, optimizing the reconstruction pipeline for small zenith angles, where image leakage is most pronounced, is vital for realizing LACT's full potential. 

{In this paper, we investigate the fundamental principles of shower reconstruction under severe image-leakage conditions using the full LACT array at a representative small zenith angle of $20^\circ$. Our analysis is organized into three methodological levels with increasing reliance on Monte Carlo simulations. First, at the single-telescope level, we use a circularized camera as a diagnostic tool to identify the origin of image-truncation bias and introduce a hybrid 2D Gaussian fitting method to improve the parameterization of severely truncated images. Second, at the stereoscopic level, we train a model on Monte Carlo data to estimate the reconstruction uncertainty of each image and use these estimates to optimally combine the telescope images through the \textit{HillasWeightedSum} and \textit{HillasWeightedDisp} methods. Finally, we explore a pixel-wise likelihood reconstruction based on Neural Ratio Estimation. Together, these studies provide a systematic evaluation of reconstruction strategies with different levels of simulation dependence, with the overall aim of optimizing LACT's angular resolution across its full field of view.}



\section{Direction Reconstruction for IACTs}
Stereoscopic direction reconstruction for IACTs is fundamentally rooted in the geometric parameterization of Cherenkov images, classically achieved via Hillas parameters. By applying Principal Component Analysis (PCA) to the image pixel distributions, one extracts the major axis of the resulting ellipse. Physically, this major axis represents the projection of the shower detector plane (the plane containing both the telescope and the shower axis) onto the camera focal plane \cite{hofmann1999comparison}. Consequently, the true source position must lie along this axis.

Building upon this geometric principle, the classical stereoscopic reconstruction determines the shower direction by calculating the weighted mean of all pairwise intersections of the major axes from multiple telescopes \cite{bernlohr2008simulation}. In this standard approach, which we refer to as \textit{HillasIntersection}, the weighting factor $w_{ij}$ for a pair of telescopes $i$ and $j$ is typically formulated to prioritize high-quality images with well-defined geometries:
$$w_{ij} = A_{amp}\sin^2(\phi_i - \phi_j) \cdot W_i \cdot W_j$$

where $A_{amp} = \frac{A_{i} A_{j}}{A_{i} + A_{j}}$ is a term scaled by the image amplitudes $A_{i}$ and $A_{j}$. The individual telescope weights are defined as $W_{i,j} = (1 - w_{i,j}/l_{i,j})$, where $w$ and $l$ denote the Hillas width and length, respectively. These weights are designed to assign higher weight to bright images with high eccentricity (elongation). Furthermore, the $\sin^2(\phi_{i} - \phi_{j})$ term ensures that pairs with large crossing angles are prioritized, thereby maximizing the effective stereoscopic reconstruction power.

However, the reconstructed major axes frequently deviate from the true source position. While inherent shower fluctuations and the optical Point Spread Function (PSF) contribute to this derivation, severe image truncation presents a far more critical structural flaw, particularly for high-altitude arrays like LACT. When a significant portion of the Cherenkov image leaks outside the camera's FoV, PCA cannot accurately capture the true moment of the distribution, leading to a skewed major axis. The severity of this deviation is quantified by the MISS parameter -- the perpendicular distance between the true source position and the reconstructed major axis -- which  serves as a direct indicator of single-telescope image quality.

To address the profound impact of image truncation and optimize the angular resolution across the entire FoV, this paper will proceed by addressing two fundamental questions:

\begin{itemize}
    \item How can we accurately evaluate and subsequently improve the image quality  for a single telescope under severe leakage conditions?

    \item How can we optimally combine images of varying qualities across the array to achieve a robust and superior stereoscopic reconstruction?
\end{itemize}

\section{Enhancing Single-Telescope Reconstruction Quality}
\begin{figure}[htbp]\centering\includegraphics[width=0.48\textwidth]{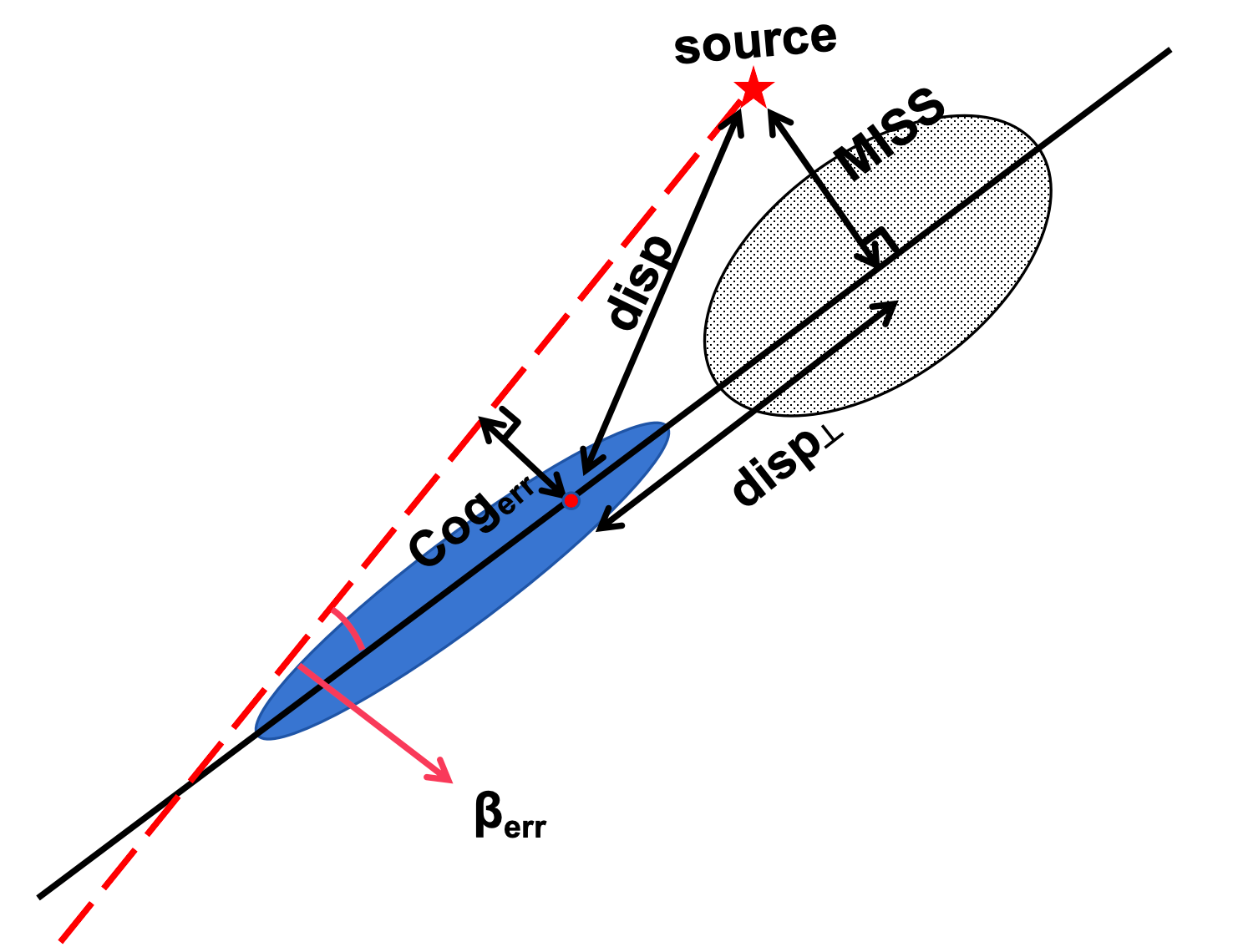}\caption{A schematic illustration of the single-telescope error components. The red line represents the true shower axis projected onto the camera plane. The blue ellipse represents the parameterized Hillas image, with its major axis indicated by the solid black line. The geometric error is decoupled into the CoG transverse displacement ($\text{Cog}_{\text{err}}$) and the angular deviation ($\beta_{\text{err}}$). The  gray ellipse indicates the error ellipse utilized in the subsequent \textit{HillasWeightedDisp} method.}\label{fig:hillas_figure}
\end{figure}

To systematically improve the overall stereoscopic angular resolution, we must first isolate and quantify the sources of geometric error inherent in a single telescope's measurement. As illustrated in Figure \ref{fig:hillas_figure}, the deviation of the reconstructed Hillas major axis from the true shower axis (projected onto the camera focal plane) can be decoupled into two distinct error components:\begin{itemize}\item {Transverse Displacement ($\text{Cog}_{\text{err}}$):} The perpendicular distance from the Center of Gravity (CoG) of the Hillas ellipse to the true shower axis. This represents a positional shift of the image centroid.\item {Angular Deviation ($\beta_{\text{err}}$):} The angular difference between the reconstructed major axis and the true shower axis. This represents a rotational skewing of the image parameterization.\end{itemize}

Following this geometric framework, and assuming a small-angle approximation, the MISS parameter can be expressed as:$$MISS \approx \text{Cog}_{\text{err}} + (\text{disp} \times \beta_{\text{err}})$$where $\text{disp}$ is the angular distance between the image CoG and the assumed source position. For a fixed impact parameter (which roughly corresponds to a fixed $\text{disp}$), the reconstruction errors $\text{Cog}_{\text{err}}$ and $\beta_{\text{err}}$ generally follow Gaussian distributions centered at zero. As their linear combination, the MISS parameter is therefore also expected to follow an approximately zero-centered Gaussian distribution. Its standard deviation, $\sigma_{\mathrm{MISS}}$, can consequently be used as a quantitative measure of the reconstruction quality of an individual telescope image.

\subsection{"Circularizing" the Camera: The Influence of Irregular Camera Shape}

The first indication of image quality degradation stems from the influence of irregular camera shapes, a factor that becomes particularly significant for events characterized by high energies and large impact parameters.

\begin{figure*}[htbp]
    \centering
    \includegraphics[width=0.48\textwidth]{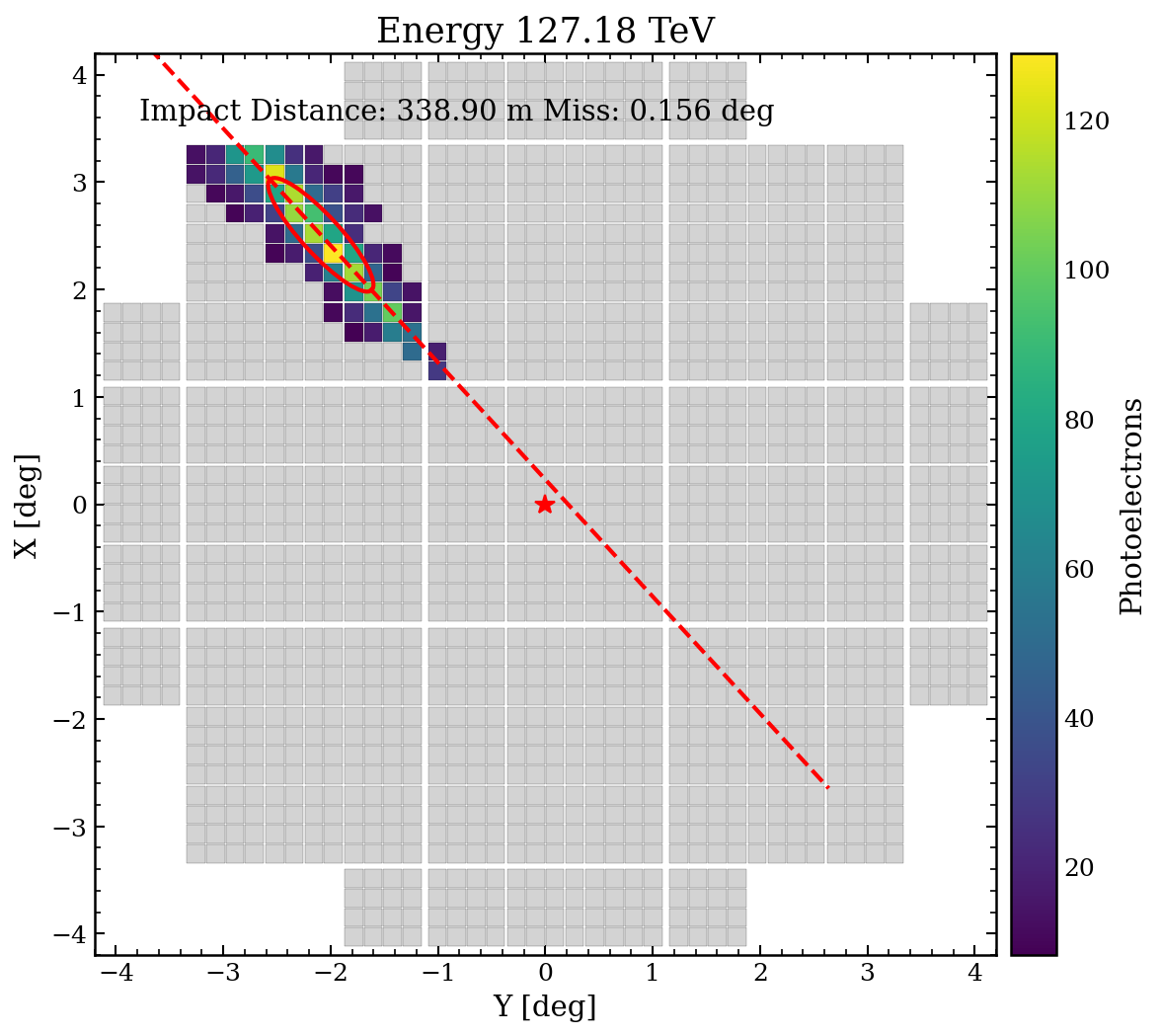}\hfill
    \includegraphics[width=0.48\textwidth]{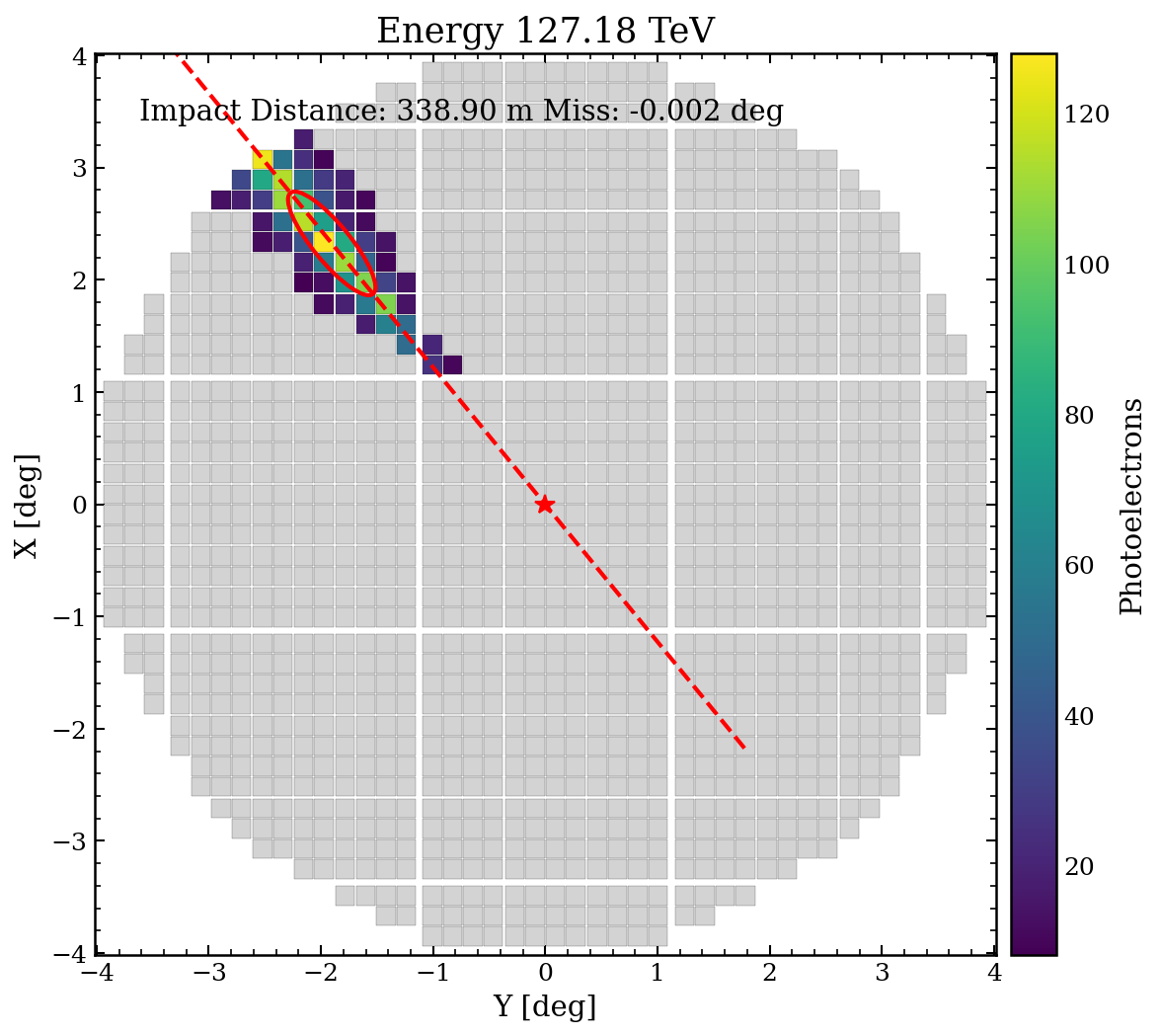}
    \caption{Comparison of the same shower event imaged by the standard LACT camera (left) and a custom circular camera cut with a $4^\circ$ radius (right). The irregular boundary of the standard camera leads to asymmetric signal leakage, skewing the major axis away from the origin. The circular cut minimizes this geometric distortion.}
    \label{fig:circular_full_camera_compare}
\end{figure*}

As illustrated in Figure \ref{fig:circular_full_camera_compare}, the same shower is imaged using both the standard LACT camera and a custom circular camera cut with a $4^\circ$ radius. It is evident that the circular camera configuration yields superior image quality. In an irregularly shaped camera, signal leakage at the boundaries asymmetrically truncates the shower image. This missing signal skews the reconstructed major axis, causing it to deviate from the true origin (which corresponds to the camera center for this particular event). In contrast, the circular camera boundary cuts the image roughly perpendicular to its major axis, resulting in minimal geometric distortion.

\begin{figure}[htbp]
    \centering
    \includegraphics[width=0.5\textwidth]{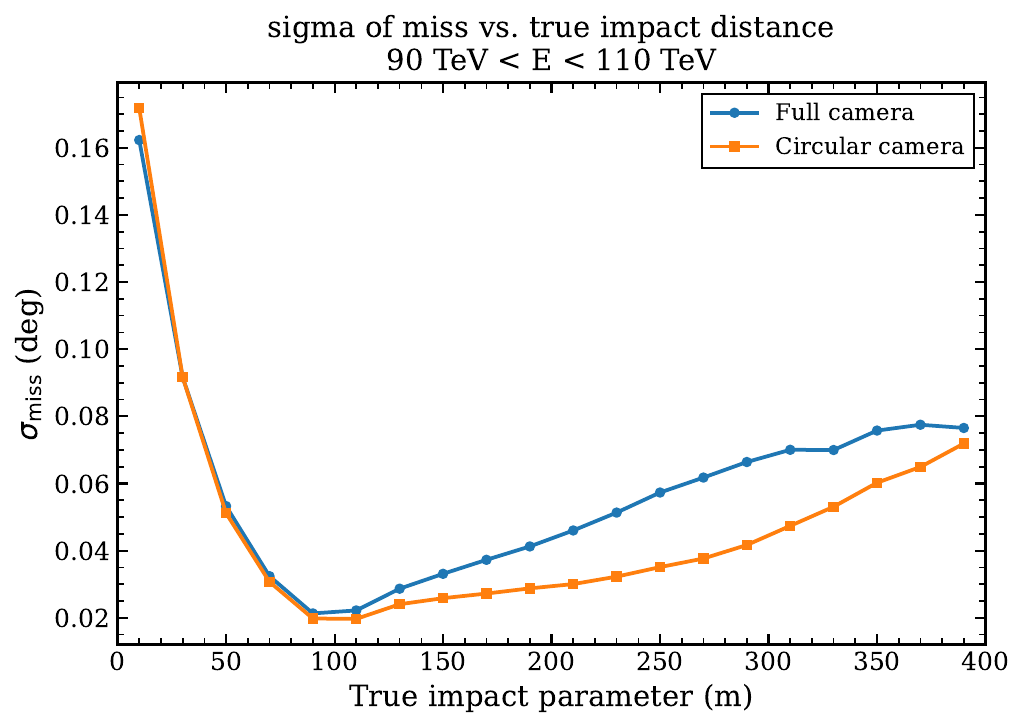}
    \caption{The standard deviation of the MISS parameter ($\sigma_{\text{MISS}}$) versus the impact parameter for events at $\sim100\text{ TeV}$. A comparison is shown between the standard full LACT camera and a custom circularized camera. The significant divergence at impact parameters greater than $100\text{ m}$ highlights the degradation caused by the irregular camera shape.}
    \label{fig:sigma_miss_versus_impact}
\end{figure}

To further quantify this effect, we examine the standard deviation of the MISS parameter ($\sigma_{\text{MISS}}$) as a function of the impact parameter for shower events at approximately $100\text{ TeV}$, as shown in Figure \ref{fig:sigma_miss_versus_impact}. At larger impact parameters (exceeding $100\text{ m}$), the discrepancy in reconstruction quality between the standard full camera and the circularized camera becomes highly significant. This divergence clearly demonstrates that the irregular camera boundary is a primary driver of the observed degradation in image quality for  large impact events.

\begin{figure}[htbp]
    \centering
    \includegraphics[width=0.48\textwidth]{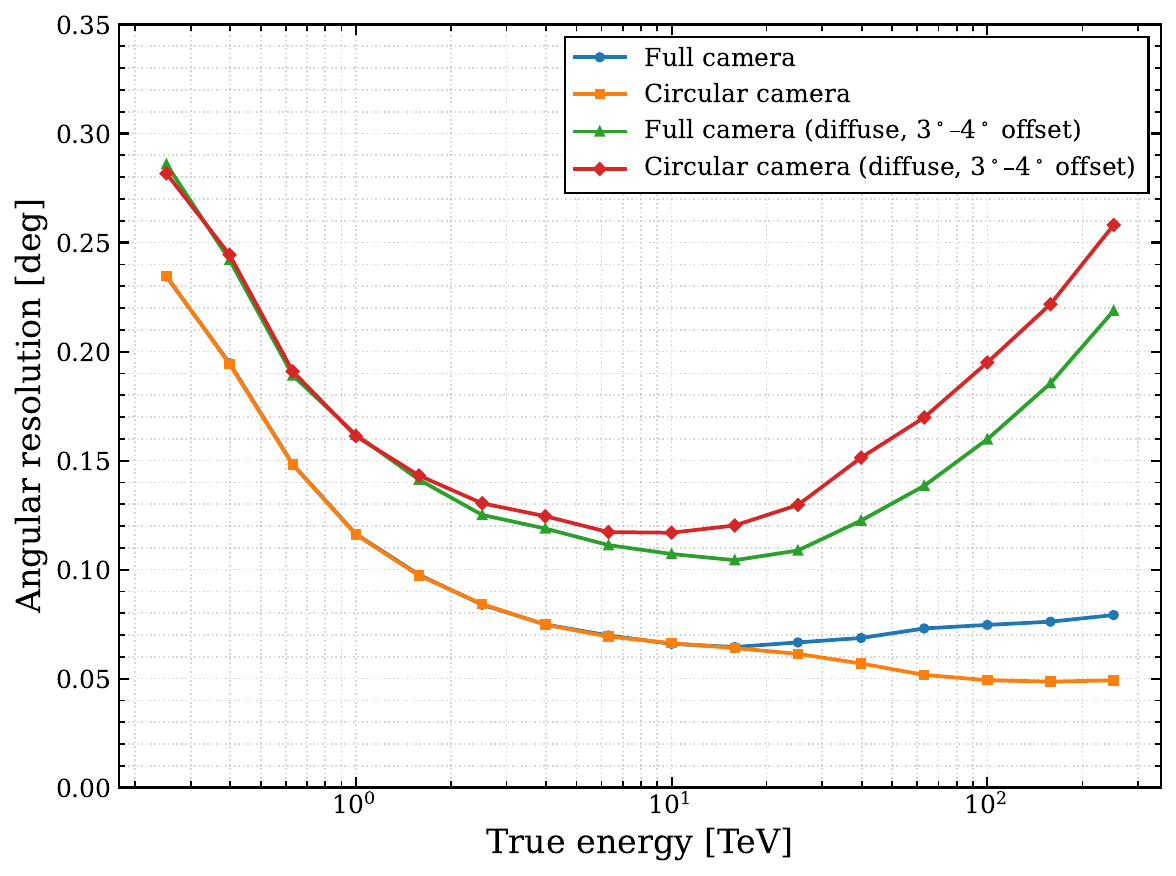}
    \caption{Comparison of the  angular resolution between the standard full LACT camera and the custom circularized camera. For on-axis sources, the circular camera shows significant improvement at high energies ($\ge 10\text{ TeV}$), achieving $\sim 0.03^\circ$ at $100\text{ TeV}$. However, for large offset sources ($3^\circ$--$4^\circ$), the circular camera performs worse due to severe truncation effects and a reduced effective field-of-view.}
    \label{fig:angular_resolution_circular_compare}
\end{figure}

Ultimately, these single-telescope geometric distortions propagate to the stereoscopic reconstruction, resulting in a significant difference in the final angular resolution. Figure \ref{fig:angular_resolution_circular_compare} compares the angular resolution achieved by both camera configurations after applying a baseline event selection, at least two telescopes passed the following condition: Intensity  $> 100$ and a Leakage2 (the fraction of signal contained in the two outermost camera pixel rings)  $< 0.3$.  {All subsequent analyses presented below adhere to these same selection criteria, ensuring they correspond to the same effective area. Also, it should be noted that the angular resolution typically improves further after gamma/hadron separation, as this background rejection step removes hadron-like events, which generally suffer from poorer reconstruction.}

For on-axis observations, the circularized camera yields a superior angular resolution above $10\text{ TeV}$. This improvement becomes increasingly pronounced at higher energies, where the leakage effect in the standard camera is most severe, reaching an optimal resolution of approximately $0.03^\circ$ at $100\text{ TeV}$.

Conversely, for events with large source offsets (e.g., $3^\circ$--$4^\circ$ off-axis), the performance of the circularized camera degrades significantly, ultimately performing worse than the standard full camera. In these large-offset scenarios, the circular boundary still suffered from  severe truncation effects. This degradation is further exacerbated by the effectively reduced field-of-view (FoV) of the circular cut compared to the full camera. 

In summary, artificially masking the camera into a circular geometry effectively mitigates the reconstruction biases introduced by an irregular camera shape. However, this mitigation is only highly effective for near-on-axis situations, where the symmetric boundary ensures a higher-quality determination of the major axis. For off-axis situations, the artificial mask reduces the effective field-of-view, resulting in even worse performance than the standard camera.

\subsection{Image Quality Degradation at Large Offset}

As illustrated in Figure \ref{fig:angular_resolution_circular_compare}, the angular resolution degrades significantly at high energies for large offset angles, worsening from $0.05^\circ$ for on-axis events to $>0.2^\circ$ at 100 TeV (which is worse than the performance of LHAASO-KM2A see appendix in \cite{cao2025ultrahigh}). This severe degradation can be attributed to several compounding factors: a decreased telescope multiplicity (the number of telescopes participating in the stereoscopic reconstruction), and the deterioration of the optical point spread function (PSF), as large-offset events predominantly illuminate the outer edge pixels of the camera where optical aberrations are more pronounced. In this section, we isolate these effects to focus specifically on the intrinsic differences in single-image quality under these large-offset conditions.

\begin{figure}[htbp]
    \centering
    \includegraphics[width=0.48\textwidth]{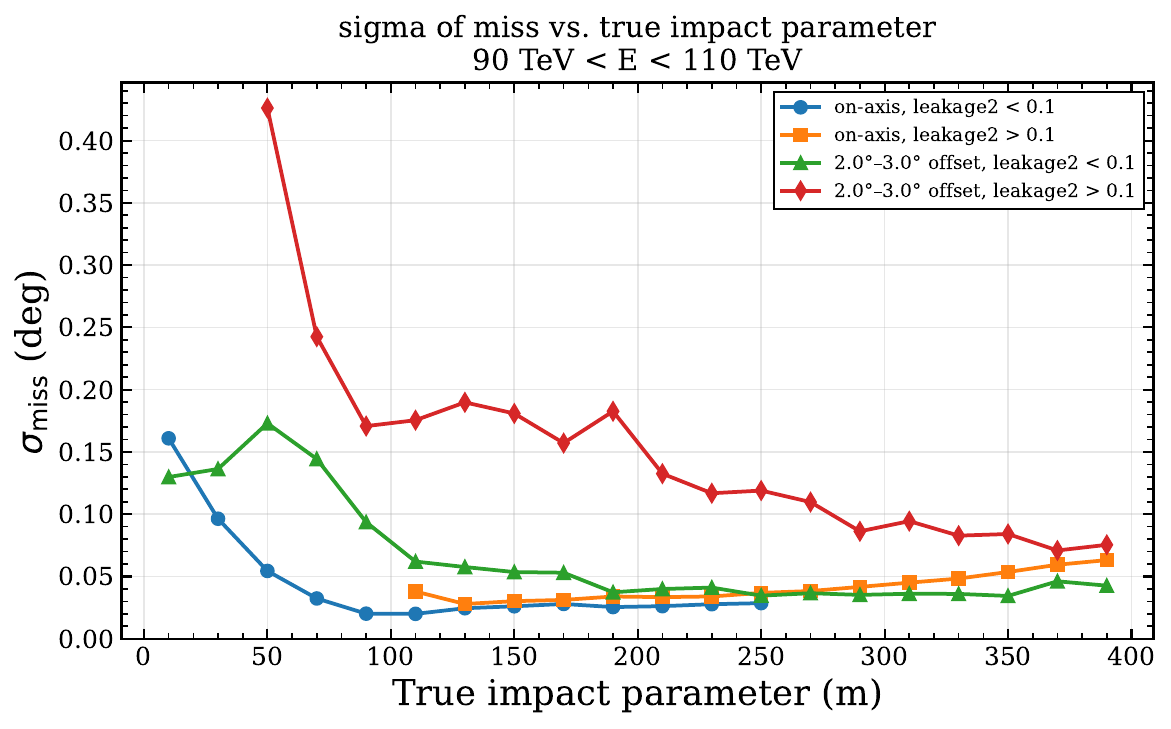}
    \caption{The standard deviation of the MISS parameter ($\sigma_{\text{MISS}}$) versus impact parameter, categorized by leakage level ($< 0.1$ and $> 0.1$). The plot compares on-axis observations with large-offset observations ($2^\circ$--$3^\circ$).}
    \label{fig:offset_miss_ip}
\end{figure}

To further investigate the degradation at large offset angles, Figure ~\ref{fig:offset_miss_ip} compares the sigma of the MISS parameter versus the impact parameter, separated by image leakage levels ($< 0.1$ and $> 0.1$). This comparison is shown for both on-axis events and off-axis events with source offsets between $2^\circ$ and $3^\circ$.

As the figure demonstrates, in the $2^\circ$--$3^\circ$ offset scenario, the discrepancy in reconstruction quality between the low-leakage ($< 0.1$) and high-leakage ($> 0.1$) samples is dominant. This divergence arises from the geometric asymmetry inherent in off-axis observations across the stereoscopic array. For a given large-offset shower, certain telescopes are favorably positioned such that the shower image falls well within their camera, effectively granting them a larger usable field-of-view (FoV) for that specific event (corresponding to a low leakage level). Interestingly, for events with large impact parameters, the images captured by these favorably positioned telescopes actually exhibit better reconstruction quality than their on-axis counterparts. 

Simultaneously, other telescopes in the array observe the same shower closer to their camera boundaries. For these telescopes, the effective FoV is severely restricted, resulting in substantial signal truncation (a high leakage level). These heavily truncated images suffer from drastically poorer quality. Ultimately, it is the inclusion of these skewed, high-leakage images in the stereoscopic analysis that drives the severe degradation of the overall angular resolution at large offset angles.

\subsection{2D Gaussian Fitting}

As established in the previous section, the primary cause of angular resolution degradation at large offset angles is the geometric truncation of shower images, which artificially skews the reconstructed major axis and deteriorates image quality. To mitigate this effect, we propose fitting the total shower image with a two-dimensional (2D) Gaussian profile. By enforcing a strict symmetric prior, this approach acts as a strong regularization technique, making the reconstruction significantly more robust against boundary truncation. 

While a similar 2D Gaussian fitting method has been successfully implemented for the VERITAS experiment \cite{christiansen2012improving}, it is even more critical for the LACT project. Operating at a high altitude and targeting very high-energy events, LACT inherently observes larger, more extended shower images that are highly susceptible to camera-edge truncation.
The 2D Gaussian spatial model is defined as:

$$
\begin{aligned}
G(x_i, y_i) = A \exp \Bigg[
&-\frac{1}{2} \left( \frac{(x_i - x_c) \cos \theta - (y_i - y_c) \sin \theta}{\sigma_{\text{length}}} \right)^2 \\
&-\frac{1}{2} \left( \frac{(x_i - x_c) \sin \theta + (y_i - y_c) \cos \theta}{\sigma_{\text{width}}} \right)^2
\Bigg]
\end{aligned}
$$
where $(x_c, y_c)$ are the coordinates of the image centroid, $\sigma_{\text{length}}$ and $\sigma_{\text{width}}$ are the standard deviations along the major and minor axes respectively, $\theta$ defines the orientation of the major axis, and $A$ is the amplitude normalization factor.To determine these six optimal parameters, we perform a $\chi^2$ minimization over the camera pixels. The $\chi^2$ function is defined as:$$\chi^2 = \sum_{i=1}^{n} \frac{\left(A_i - G(x_i, y_i)\right)^2}{G(x_i, y_i) + \sigma_{\text{ped}}^2}$$where $A_i$ is the measured signal intensity in the $i$-th pixel, $n$ is the total number of pixels included in the fit, and the denominator represents the total variance for each pixel. 

\begin{figure}[htbp]
    \centering
    \includegraphics[width=0.48\textwidth]{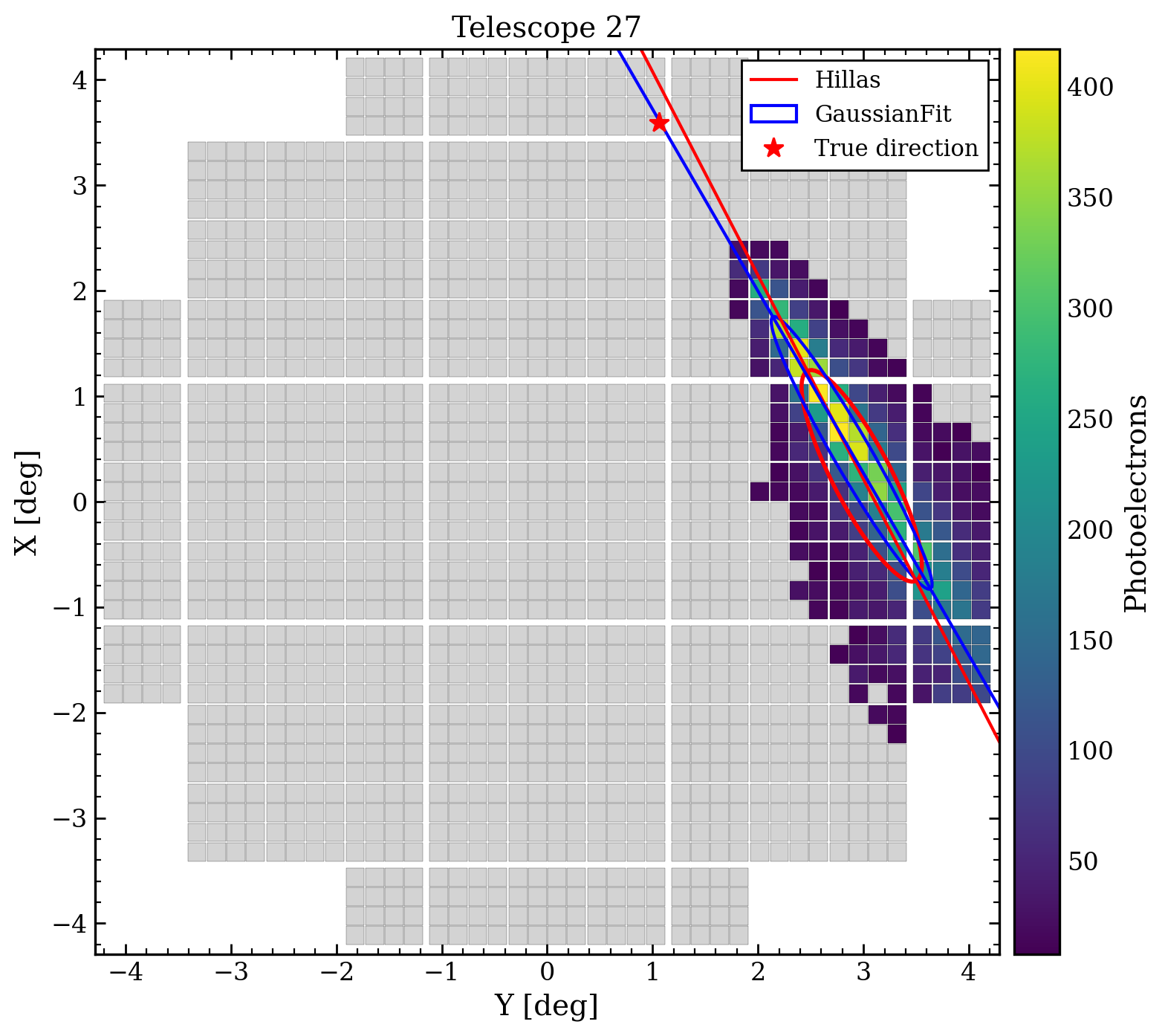}
    \caption{A truncated shower image at the camera edge, comparing the standard Hillas parameterization  with the 2D Gaussian fit.}
    \label{fig:gaussian_camera_image}
\end{figure}

A comprehensive comparison between the 2D Gaussian fitting method and standard Hillas parameterization has been conducted. We find that for faint and well-contained images, Hillas parameterization is not only significantly faster to compute, but it also yields superior image quality, demonstrating its inherent robustness in these regimes. However, for bright but truncated images, the 2D Gaussian fitting approach is heavily favored, providing substantially better reconstruction quality. Figure \ref{fig:gaussian_camera_image} displays a truncated shower image overlaid with both the standard Hillas ellipse and the fitted 2D Gaussian ellipse, visually illustrating the advantage of the spatial fitting method in recovering the true shower geometry. 

\begin{figure}[htbp]
    \centering
    \includegraphics[width=0.48\textwidth]{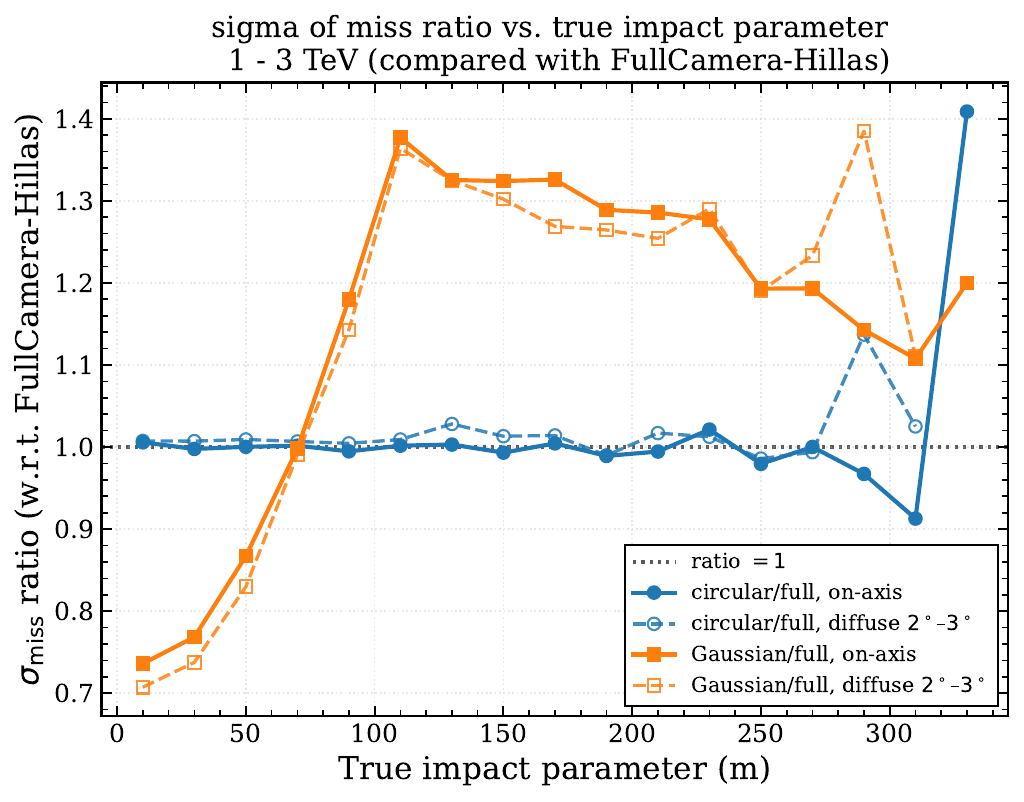}\hfill
    \includegraphics[width=0.48\textwidth]{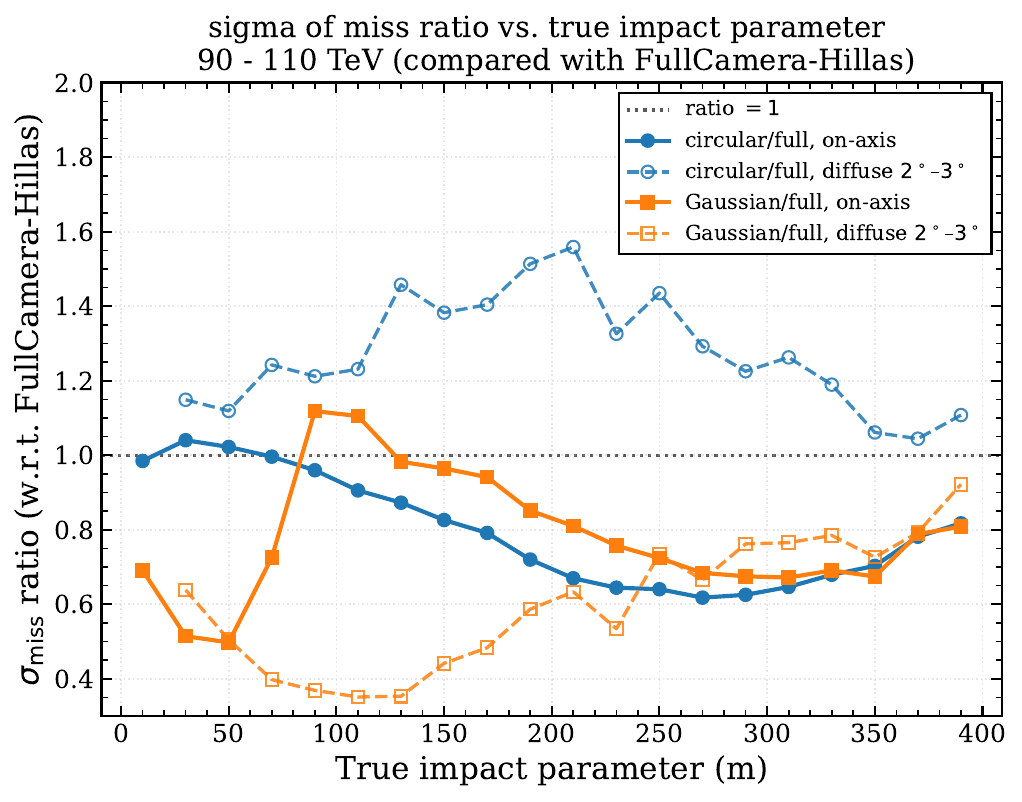}
    \caption{Comparison of the MISS parameter ratio versus impact parameter for low energies (1--3 TeV, left) and high energies (90--110 TeV, right). The 2D Gaussian fitting method demonstrates consistent robustness across different offset angles, avoiding the severe off-axis degradation seen in the circularization method.}
    \label{fig:miss_vs_ip_ratio}
\end{figure}

More importantly, as shown in Figure \ref{fig:miss_vs_ip_ratio}, unlike the camera circularization technique (which remains the optimal method strictly for on-axis observations), the 2D Gaussian fitting method demonstrates strong robustness across a wide range of source offset angles. 

To systematically incorporate the advantages of both techniques, we adopt a hybrid approach. The 2D Gaussian fit results are selectively utilized only when the following two criteria are met:
\begin{enumerate}
    \item \textbf{High Leakage ($\text{leakage2} > 0.1$):} This ensures the fit is only applied when edge truncation actively degrades the standard parameterization.
    \item \textbf{Sufficient Signal Density:} The ratio of the total image intensity to the fitted intensity ($A_{\text{all}} = 2\pi A \sigma_{\text{width}}  \\ \sigma_{\text{length}}$) must be greater than $0.2$. This condition guarantees that a sufficient signal is present to reliably constrain the six parameters of the 2D fit.
\end{enumerate}

\begin{figure}[htbp]
\centering\includegraphics[width=0.48\textwidth]{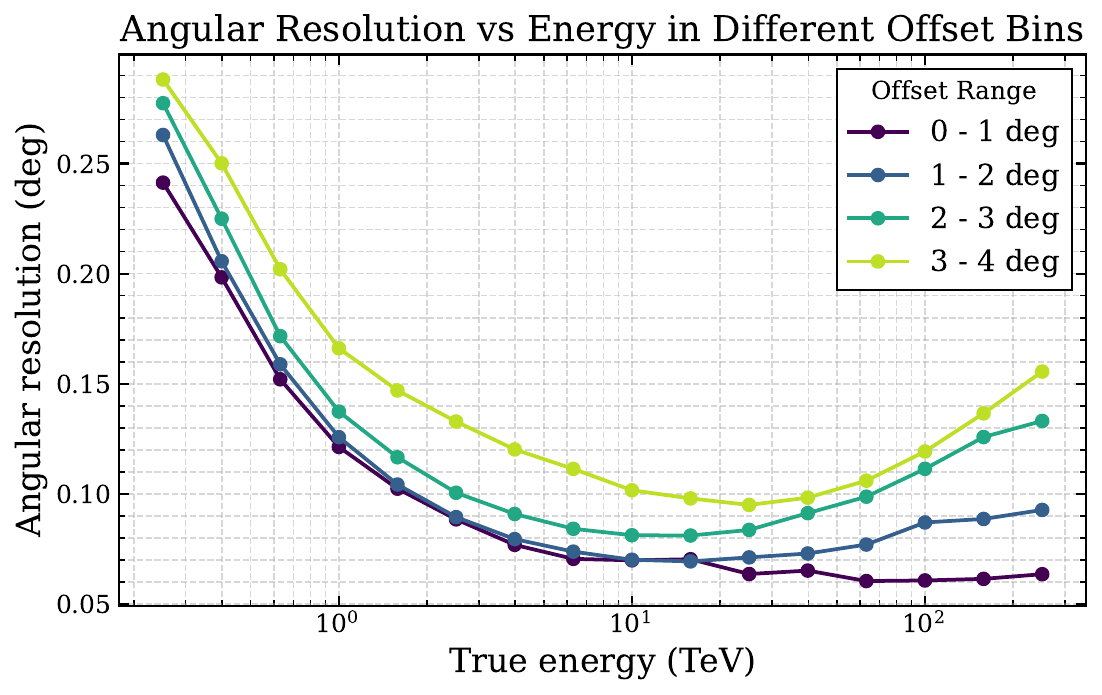}
\caption{The angular resolution(defined as 68\% conatinment of direction error) as a function of energy for different source offset bins, utilizing the optimal hybrid baseline method (combining Hillas parameterization and 2D Gaussian fitting).}

\label{fig:angular_resolution_gaussian}
\end{figure}

{By applying this hybrid approach, we establish an optimal baseline for further analysis. Utilizing the two selection criteria defined above, we evaluate the resulting stereoscopic angular resolution across various source offset bins, as presented in Figure \ref{fig:angular_resolution_gaussian}. The hybrid method yields a significantly flatter, more consistent angular resolution across the entire field of view. In the $0^{\circ}$–$1^{\circ}$ offset bin, the angular resolution is better than $0.06^{\circ}$ at $100\text{ TeV}$, outperforming even the circular camera case. Across the entire field of view, the resolution remains better than $0.12^{\circ}$ at this energy scale. Crucially, an improvement of 25\%–30\% is achieved in the $3^{\circ}$–$4^{\circ}$ offset bin, marking a substantial advancement over the standard Hillas parameterization and effectively mitigating the severe off-axis degradation caused by image truncation.}

\section{Combining Images of Varying Quality}

Having improved the image quality at the single-telescope level, the challenge now shifts to elegantly combining images of varying quality across the stereoscopic array. The classical \textit{HillasIntersection} method relies on a pairwise combination of telescope axes. While it incorporates weights based on intensity and intersection angles, this pairwise approach has a fundamental limitation: a high-quality telescope (e.g., a bright, nearby image) is often paired with a low-quality telescope (e.g., a dim, distant image). The resulting intersection point inherits the uncertainty of the poorer telescope, degrading the final weighted average.

To overcome this limitation and further advance reconstruction performance, it is necessary to decouple these pairs and evaluate each telescope independently. By doing so, we can assign statistical weights based purely on individual image quality, ensuring that superior measurements dominate the reconstruction. In this section, we first discuss methods for quantitatively estimating single-image quality. Subsequently, we implement the reconstruction algorithm introduced by \cite{hofmann1999comparison} to optimally weight these independent measurements and enhance the overall  angular resolution.

\subsection{Estimating Single-Image Quality}
\label{chap:single_quality}
As previously discussed, the overall image quality can be decoupled into two distinct error components, $\mathrm{Cog}_{\mathrm{err}}$ and $\beta_{\mathrm{err}}$, both of which are assumed to follow a zero-centered Gaussian distribution. Consequently, estimating the standard deviation ($\sigma$) of these distributions provides a quantitative measure of single-image quality. Traditionally, as seen in \cite{hofmann1999comparison,lu2013improving,stamatescu2010new}, these uncertainties are parameterized using look-up tables or analytical functions that primarily rely on basic Hillas parameters, such as image intensity and the width-to-length ratio.

To improve upon these classical techniques, we implement a machine learning approach utilizing quantile regression via the LightGBM\cite{ke2017lightgbm} framework. Rather than predicting a single mean value, this method directly predicts specific quantiles of the error distributions based on a multidimensional set of input features. For each offset bin, we train two independent quantile regression models to predict the $16^{\mathrm{th}}$ ($Q_{16}$) and $84^{\mathrm{th}}$ ($Q_{84}$) percentiles for both $\beta_{\mathrm{err}}$ and $\mathrm{cog}_{\mathrm{err}}$. From these predictions, the effective standard deviation is derived as:

$$\sigma = \frac{Q_{84} - Q_{16}}{2}$$

\begin{figure*}[htbp]
    \centering
    \includegraphics[width=0.8\textwidth]{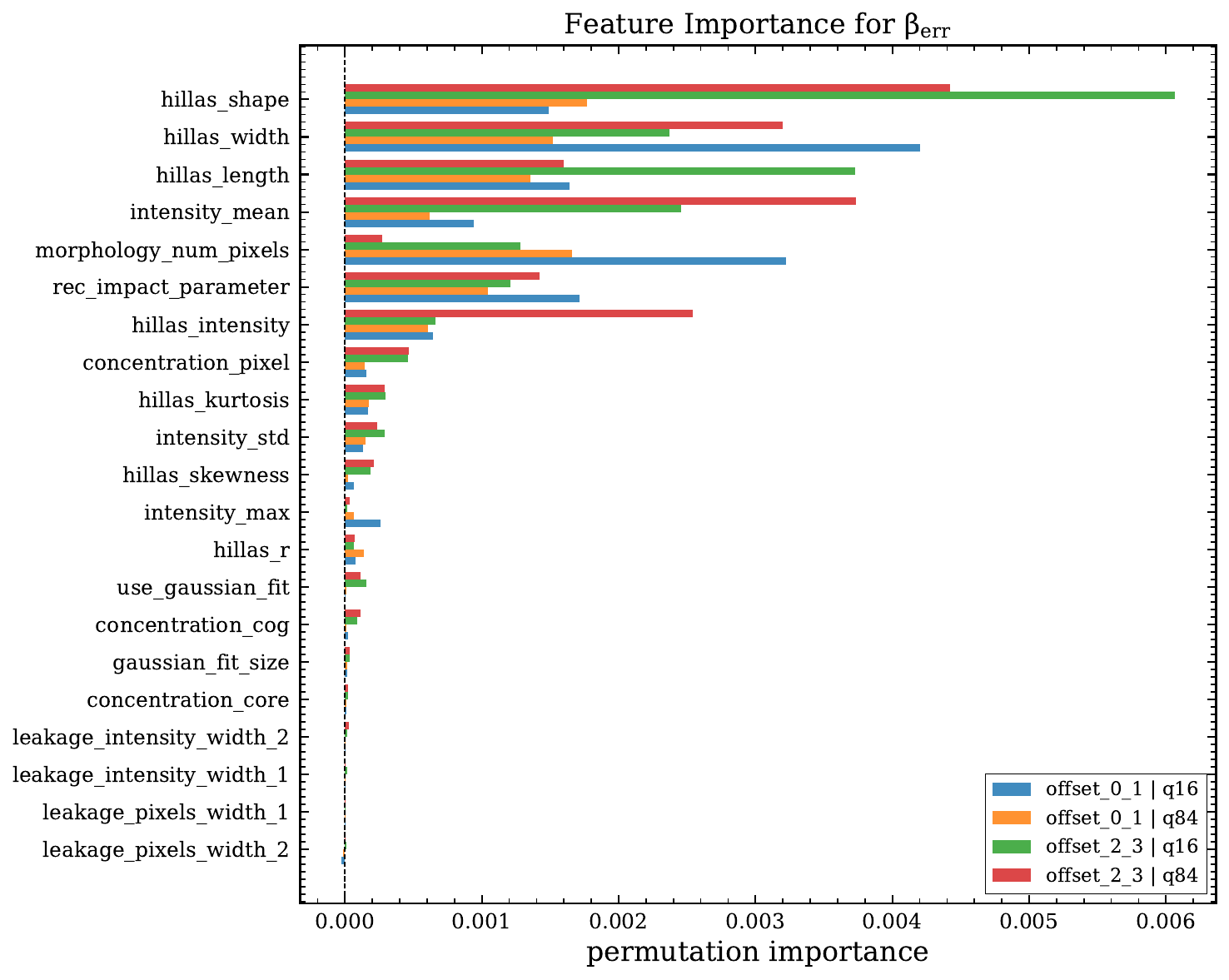}
    \caption{Permutation importance of the input features used in the LightGBM quantile regression model to predict $\beta_{\mathrm{err}}$. The importance is compared across two different offset bins ($0^\circ-1^\circ$ and $2^\circ-3^\circ$) for the $16^{\mathrm{th}}$ and $84^{\mathrm{th}}$ percentile predictions.}
    \label{fig:feature_importance_beta}
\end{figure*}

The permutation importance of the input features used to predict $\beta_{\mathrm{err}}$ is illustrated in Figure~\ref{fig:feature_importance_beta} for two distinct offset bins. As anticipated, the image intensity and the Hillas shape parameter (defined as the width-to-length ratio) emerge as quite important features. All input parameters follow the standard definitions established in the \textit{ctapipe} framework \cite{linhoff2024ctapipe}, with comprehensive descriptions provided in Appendix ~\ref{app:parameter}. A notable takeaway from this analysis is the negligible influence of the leakage parameter. This indicates that our Gaussian fitting method successfully mitigates the classical leakage problem, rendering the reconstructed image quality largely independent of the degree of shower truncation at the camera edge.

\begin{figure*}[htbp]
    \centering
    \includegraphics[width=0.8\textwidth]{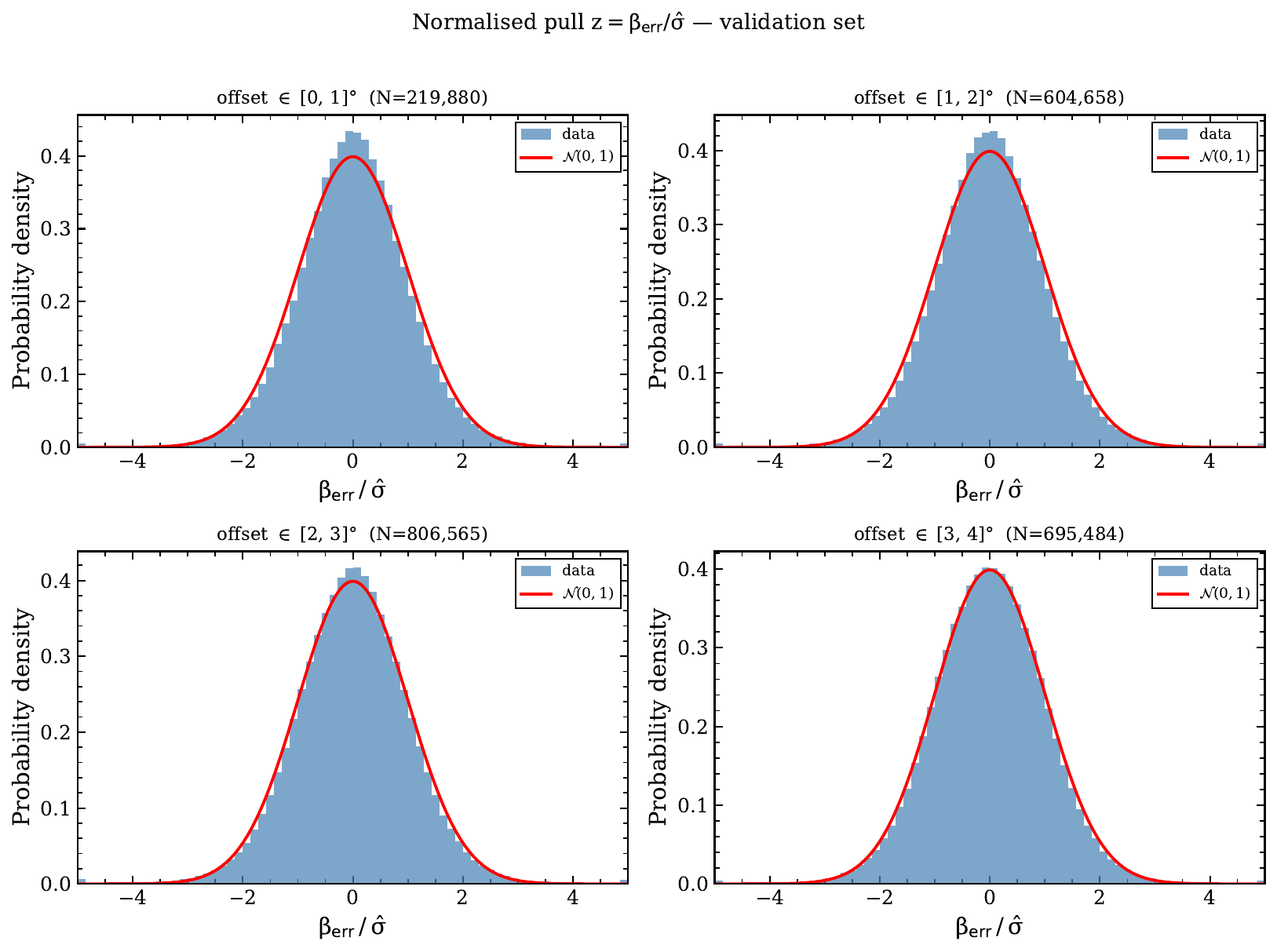}
    \caption{Normalized pull distributions ($\beta_{\mathrm{err}}/\sigma_{\beta}$) evaluated on an independent validation dataset. The distributions are shown across different offset bins and demonstrate strong agreement with a standard normal distribution, $\mathcal{N}(0,1)$, verifying the accuracy of the estimated uncertainties.}
    \label{fig:normalized_pull_beta}
\end{figure*}

To validate the accuracy of our error estimation, we calculate the normalized pull, defined as $\beta_{\mathrm{err}}/\sigma_{\beta}$, using an independent validation dataset. As illustrated in Figure~\ref{fig:normalized_pull_beta}, the resulting pull distributions for each offset bin closely follow a standard normal distribution, $\mathcal{N}(0,1)$. This strong agreement verifies that our estimated uncertainties accurately reflect the true reconstruction errors, thereby demonstrating the robustness of our quantile regression approach.

\subsection{\textit{HillasWeightedSum} Method}
As a primary approach, we implemented Algorithm 2 from \cite{hofmann1999comparison}, which we hereafter refer to as the \textit{HillasWeightedSum} method. This technique determines the optimal source position through a global minimization of the perpendicular distances between a candidate source coordinate and the major axes of the images from all telescopes. The objective function to be minimized is defined as the weighted sum of these squared distances:$$\chi^2 (x, y) = \sum_{i=1}^{N_{\mathrm{tel}}} \left( \frac{d_i(x, y)}{\sigma_{\mathrm{miss}, i}} \right)^2$$where $(x, y)$ denotes the candidate source coordinates in the camera plane, $d_i$ represents the perpendicular distance from these coordinates to the major axis of the $i$-th telescope, and the statistical weight is given by the inverse square of the estimated uncertainty, $\sigma_{\mathrm{miss}, i}$. An analytical solution for this minimization problem can be explicitly derived, the details of which are provided in Appendix~\ref{app1}.

Figure~\ref{fig:perf_weighted_sum} illustrates the angular resolution achieved by the \textit{HillasWeightedSum} method compared to the classical \textit{HillasIntersection} approach. To establish a theoretical upper bound on performance, we also include the results of the \textit{HillasWeightedSum} method using the true MISS parameters as weights. Overall, a modest but consistent improvement is observed across all offset bins.

At lower energies ($E < 1\,\mathrm{TeV}$), the performance of both methods converges, and nearlly no improvement is seen even at the theoretical limit. This behavior is expected, as low-energy showers frequently trigger only two telescopes (multiplicity $N=2$), a scenario in which the \textit{HillasWeightedSum} analytical solution naturally reduces to the simple geometric intersection point of the \textit{HillasIntersection} method. Conversely, for high-energy events at large offsets, the \textit{HillasWeightedSum} algorithm yields a more substantial improvement of approximately $0.02^\circ$ to $0.03^\circ$. This performance gain clearly demonstrates the efficacy of our independent weighting scheme. Furthermore, it highlights that under such conditions, the classical pairwise weighting of the \textit{HillasIntersection} method is suboptimal and benefits significantly from the inclusion of single-image quality metrics.

\begin{figure*}[htbp]
    \centering
    \includegraphics[width=0.8\textwidth]{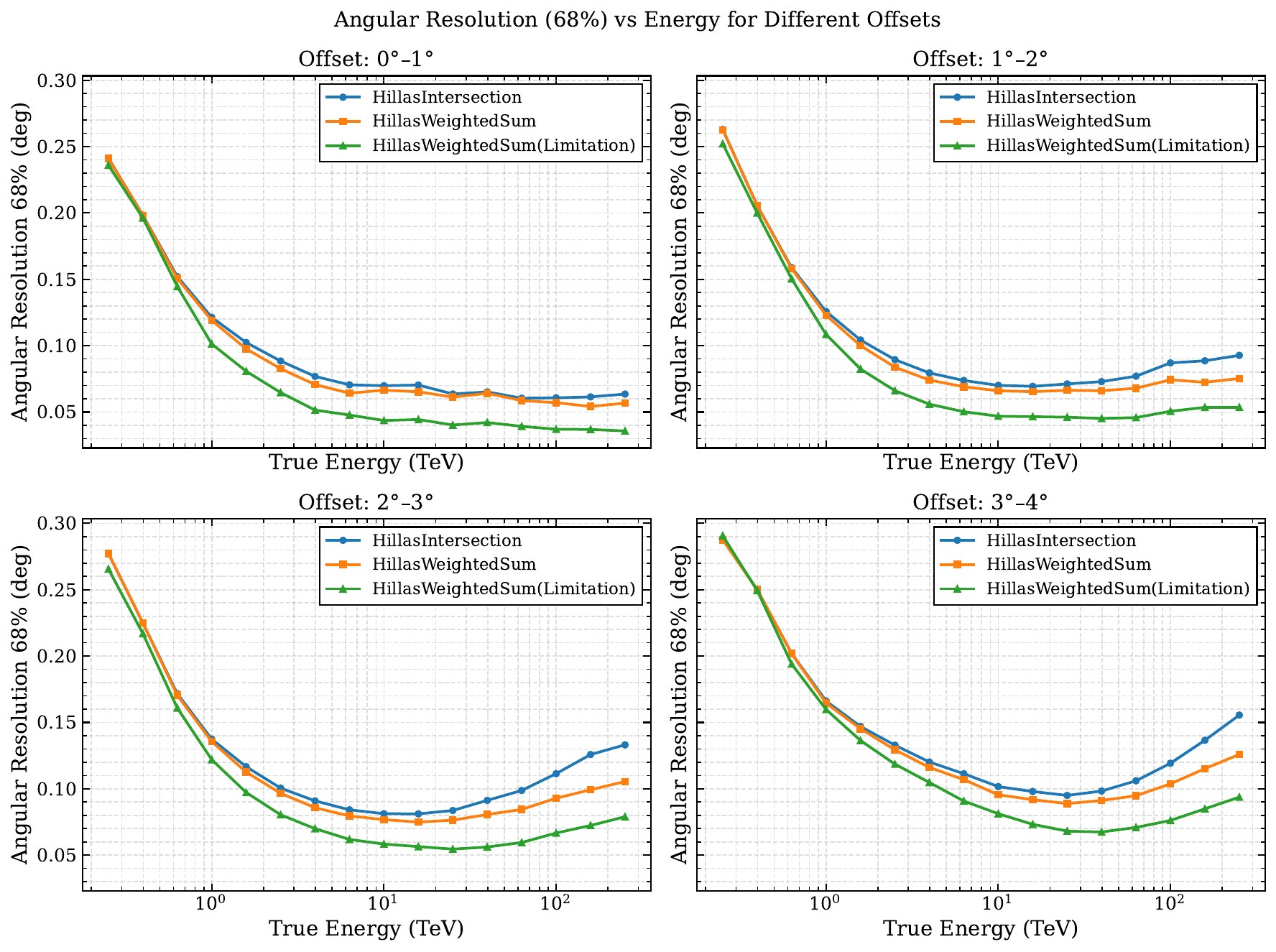}
    \caption{Comparison of the reconstruction performance between the \textit{HillasWeightedSum} and classical \textit{HillasIntersection} methods across different offset bins. The theoretical limit of the \textit{HillasWeightedSum} approach (using true \textit{miss} values for weighting) is also shown to illustrate the maximum achievable improvement.}
    \label{fig:perf_weighted_sum}
\end{figure*}

\subsection{\textit{HillasWeightedDisp} Method}
The Disp method was originally proposed for single-telescope reconstruction by the Whipple collaboration \cite{fomin1994new, lessard2001new} and has since been widely adopted by other experiments\cite{albert2008vhe, abe2023observations}.

The core principle of the Disp method is that the shape of the shower image encodes information regarding the distance between the image's Center of Gravity (COG) and the source position.

Beyond its application in single-telescope analysis, the Disp method is also highly effective for stereoscopic reconstruction. It is particularly robust against poor image quality, specifically in scenarios involving large offset angles\cite{lypova2021analysis} or nearly parallel images, which often occur during large zenith angle observations \cite{lu2013improving, csenturk2011disp}. 

This approach corresponds to Algorithm 3 described by Hofmann et al. \cite{hofmann1999comparison} (hereafter referred to as the \textit{HillasWeightedDisp} method). In this framework, each telescope naturally provides an independent estimate of the source position, which is then combined via a weighted average. Since the reconstruction requires at least two telescopes, the head-tail ambiguity\cite{domingo2005disp}  is avoided entirely. For simplicity, we always select the direction closer to the initial value provided by  \textit{HillasIntersection} method.

In the \textit{HillasWeightedDisp} framework, besides the uncertainty mentioned in Section \ref{chap:single_quality}, it is necessary to estimate both the Disp value and its corresponding uncertainty. Here, each telescope naturally yields an independent estimation and an associated error ellipse; the analytical optimal solution for the weighted average is presented in Appendix \ref{app2}.

\begin{figure*}[htbp]
    \centering
    \includegraphics[width=0.8\textwidth]{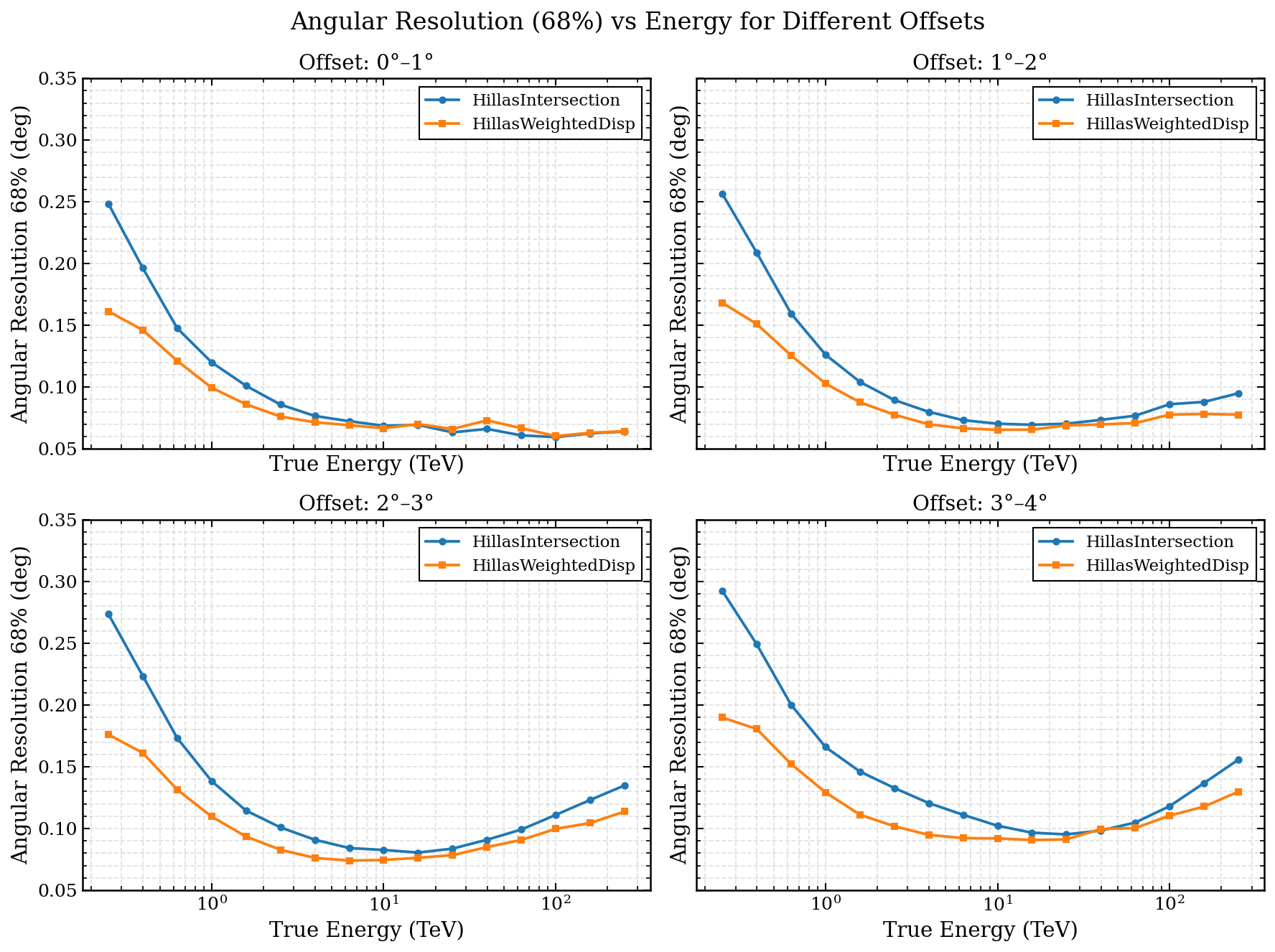}
    \caption{Comparison of reconstruction performance between the \textit{HillasWeightedDisp} and \textit{HillasIntersection} methods.}
    \label{fig:weighted_disp_perf}
\end{figure*}

The performance comparison with the \textit{HillasIntersection} method is presented in Figure~\ref{fig:weighted_disp_perf}.
In the high-energy regime, the method exhibits a trend consistent with the \textit{HillasWeightedSum} approach: the performance improvement becomes increasingly pronounced at higher energies and larger offset angles. Moreover, the \textit{HillasWeightedDisp} method yields a substantial improvement in the low-energy regime ($< 1\,\mathrm{TeV}$). Unlike the standard intersection method, which relies heavily on the precise determination of the image's major axis (a parameter that is often poorly constrained for faint, low-intensity images), the Disp method leverages the overall image shape and centroid information. Consequently, it is significantly less sensitive to the poor image quality characteristic of low-energy events, enabling more accurate reconstructions in scenarios where the traditional intersection method typically fails.

\textbf{\section{Exploring Pixel-wise Likelihood Reconstruction}}

In reality, Cherenkov images exhibit intrinsic asymmetry, particularly along the major axis due to the longitudinal development of the shower (commonly characterized by the skewness). Recognizing this limitation, a more rigorous, first-principles approach naturally shifts away from a global 2D Gaussian assumption toward evaluating the expected charge distribution on a pixel-by-pixel basis. By doing so, we can directly define the probability density for each pixel given a specific set of shower parameters, denoted as $P(q_i | \mathbf{X})$. Here, $q_i$ is the observed pixel charge and $\mathbf{X}$ represents the shower state vector, including the direction, core position, energy, and depth of maximum ($X_{\mathrm{max}}$). Assuming the pixel measurements are independent, the optimal shower parameters can be extracted via Maximum Likelihood Estimation by minimizing the negative log-likelihood function:
$$-\ln \mathcal{L}(\mathbf{X}) = -\sum_{i} \ln P(q_i | \mathbf{X})$$


This concept forms the foundation of advanced reconstruction algorithms developed by the CAT \cite{le1998new} and H.E.S.S. \cite{de2009high} collaborations. In \cite{de2009high}, the authors parameterized the air shower to predict the mean charge value for each pixel under specified shower parameters, utilizing an analytical function that accounts for the pedestal and electronic noise to describe the pixel-wise probability distribution. Later, the ImPACT method \cite{parsons2014monte} extended this approach by using real Monte Carlo simulations to generate expected templates. This avoids the analytical approximations that struggle with large fluctuations in high-energy events, thereby achieving better performance and becoming widely used in scientific publications \cite{abdalla2018hess, hess2024acceleration} .

More recently, the FreePACT method \cite{schwefer2024hybrid} improved upon template-based approaches by leveraging deep learning to maximize the likelihood directly. This eliminates the need for an analytical formula and significantly reduces the computational time required to interpolate between templates, further enhancing reconstruction performance and demonstrating the immense potential of this technique. Inspired by these advancements, we adopt a strategy similar to FreePACT to validate its potential for the LACT array, with a particular focus on large-offset events.

\begin{figure*}[htbp]
    \centering
    \includegraphics[width=0.48\textwidth]{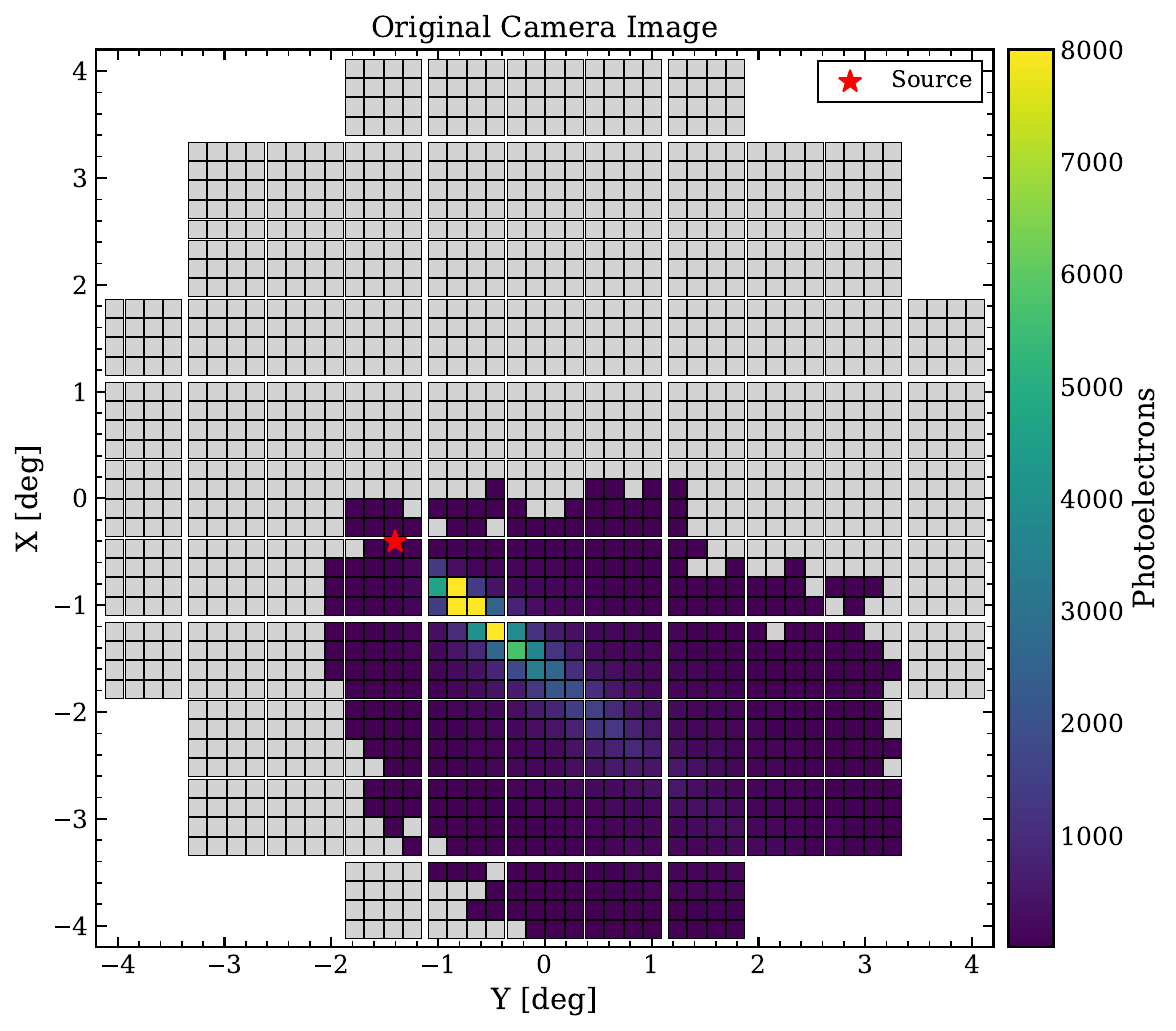}
    \hfill
    \includegraphics[width=0.48\textwidth]{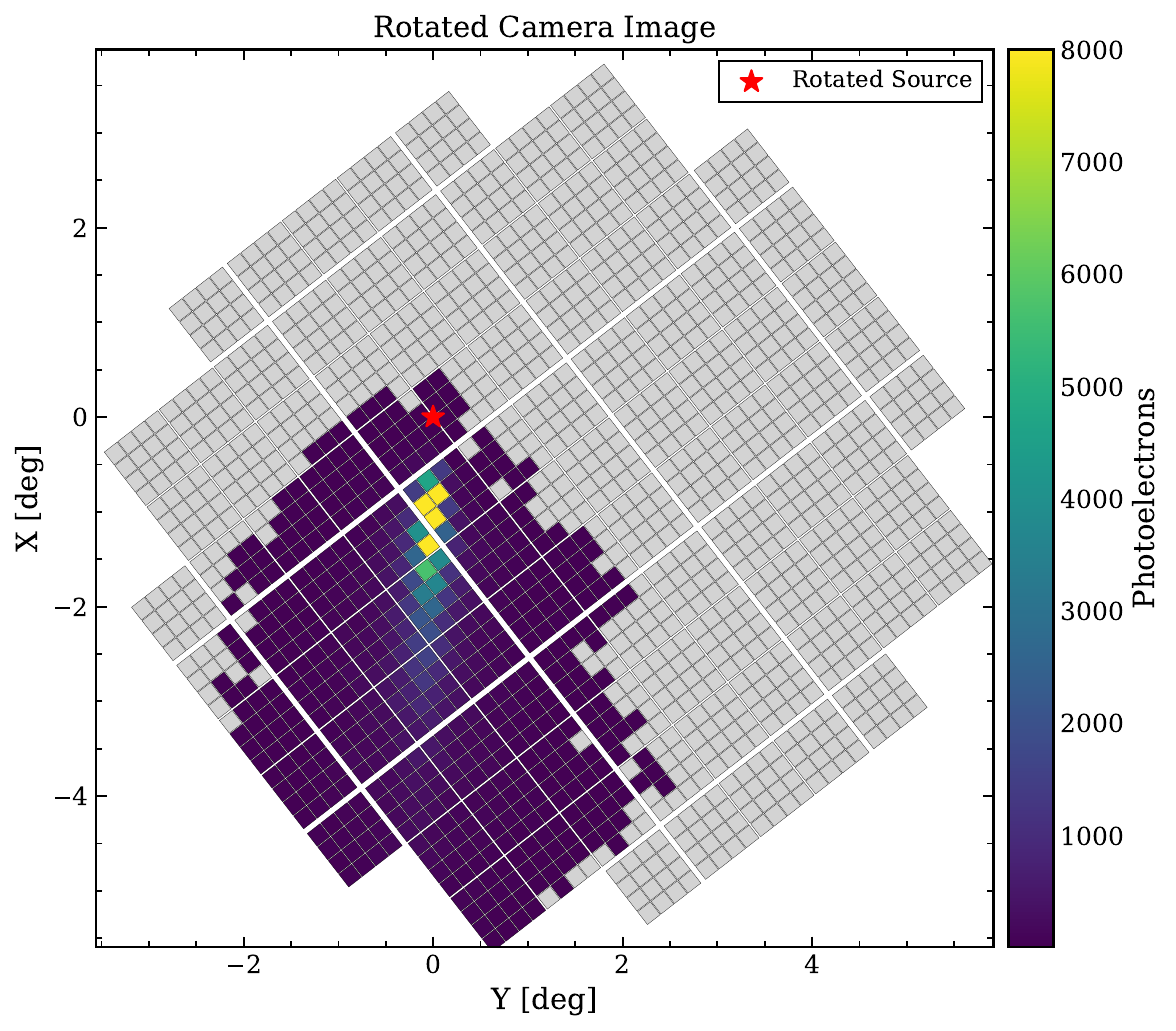}
    \caption{Comparison of the camera images before (left) and after (right) the coordinate transformation. The transformed data centers the source position and aligns the image along the x-axis to exploit physical symmetries.}
    \label{fig:camera_image_compare}
\end{figure*}

 We adopt a transformation similar to that described in \cite{schwefer2024hybrid} to fully leverage physical symmetries and reduce problem complexity. Exploiting the approximate symmetry in the camera plane, we transform the pixels into a coordinate system where the source position lies at the center of the camera, and the images are rotated to align with the x-axis. This transformation process is illustrated in Figure~\ref{fig:camera_image_compare}. The subsequent analysis utilizes these transformed pixel coordinates, $(x^{\prime}, y^{\prime})$, along with their corresponding charges. To derive the probabilities, we employ a similar Neural Ratio Estimation \cite{hermans2020likelihood} approach. Specifically, a classifier is trained to differentiate between genuine and shuffled combinations of the shower parameters (depth of maximum $X_{\mathrm{max}}$, energy $E$, and impact parameter), pixel positions, and charges.

In contrast to FreePACT, we train our model using a diffuse gamma-ray dataset. This approach provides two distinct advantages. First, it inherently accounts for varying optical qualities across the telescope's field of view. Second, it supplies necessary training samples for transformed pixel positions that fall outside the physical camera boundaries; because our coordinate transformation centers the source position, off-axis events naturally map to transformed coordinates that are absent in purely on-axis training sets. Furthermore, for simplicity, we restrict our analysis to pixels that survive standard tail-cut cleaning. 

\begin{figure*}[htbp]
    \centering
    \includegraphics[width=0.8\textwidth]{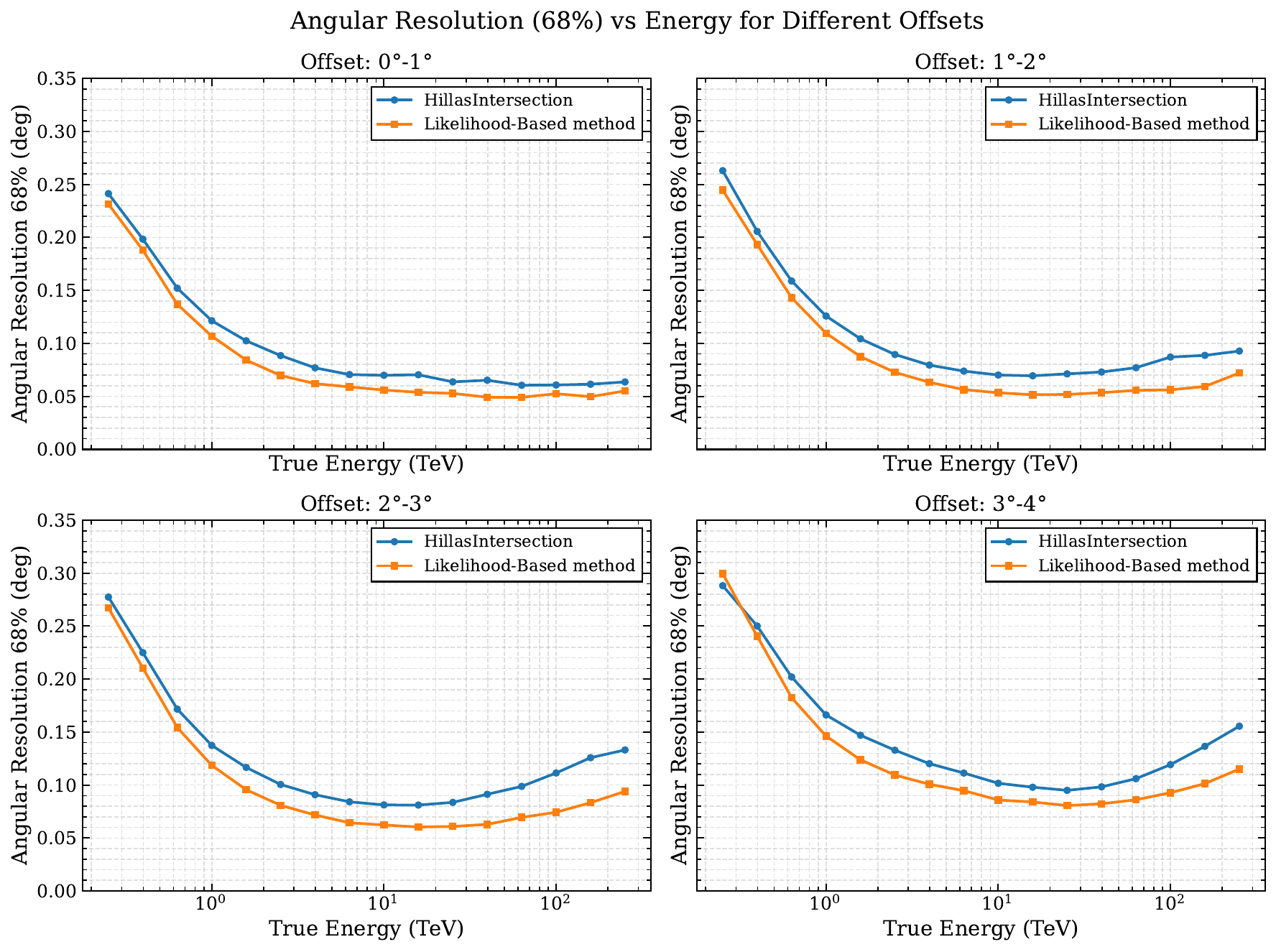}
    \caption{The  performance of the likelihood-based model across different offset bins. \textbf{The performance  here is obtained under the nominal Monte Carlo simulations.}}
    \label{fig:freepact_perf}
\end{figure*}

{The final performance across different offset bins is shown in Figure~\ref{fig:freepact_perf}. By exploiting pixel-level information, this method significantly outperforms the standard Hillas-based analysis, with the improvement becoming increasingly pronounced at larger offset angles. Within a $3^{\circ}$ offset, the angular resolution at $100~\rm TeV$ remains highly stable between $0.05^{\circ}$ and $0.07^{\circ}$, and is better than $0.09^{\circ}$ across all offset bins. Notably, in the $2^{\circ}$--$3^{\circ}$ offset bin, the improvement at $100~\rm TeV$ reaches approximately 35\% , representing a substantial gain in reconstruction precision.}

Despite these promising results, the efficacy of likelihood-based methods is fundamentally constrained by the precision of the underlying Monte Carlo simulations. The "sim-to-real gap"  must be thoroughly minimized to ensure agreement between simulations and observations. \textbf{Consequently, we regard this study as an exploratory attempt to estimate the potential improvement in LACT’s angular resolution offered by likelihood-based reconstruction under nominal simulation conditions.} This reliance on ideal simulations also dictates our decision to defer the deployment of fully data-driven deep learning models (e.g., CTLearn \cite{nieto2019ctlearn}) until they can be rigorously validated against real prototype observations \cite{zhang2026lact}.

\section{Dicussion}
\begin{table*}[t]
\centering
\caption{
Summary of the reconstruction methods investigated in this work.
The improvement is evaluated at 100~TeV for the representative offset range
of $2^\circ$--$3^\circ$. The last column shows the corresponding improvement
for simulations with increased night-sky background (NSB).
}
\label{tab:method_comparison}

\small
\setlength{\tabcolsep}{4pt}

\begin{tabularx}{\textwidth}{
    @{}
    p{0.17\textwidth}
    X
    p{0.16\textwidth}
    p{0.19\textwidth}
    p{0.19\textwidth}
    @{}
}
\toprule
Method
& Main feature
& Comparison baseline
& Improvement 
& Improvement at increased NSB \\
\midrule

Hybrid Gaussian fitting
& Robust reconstruction of strongly truncated images
& Classical Hillas
& $\sim25\%$ ($\sim0.035^\circ$)
& ------ \\

Uncertainty-weighted reconstruction
& Accounts for different reconstruction quality among telescopes
& Hybrid Gaussian
& $\sim18\%$ ($\sim0.020^\circ$)\newline
  (\textit{HillasWeightedSum})
& $17\%$ at 200\% NSB\newline
  $14\%$ at 300\% NSB \\

Pixel-wise likelihood reconstruction
& Exploits full pixel-level information
& Hybrid Gaussian
& $\sim35\%$ ($\sim0.038^\circ$)
& $34\%$ at 200\% NSB\newline
  $31\%$ at 300\% NSB \\

\bottomrule
\end{tabularx}
\end{table*}
\subsection{Simulation dependence and robustness}
{An important limitation of the present study is that the performance evaluation is based entirely on Monte Carlo simulations. Actually, the reconstruction methods investigated here have different levels of dependence on the accuracy of these simulations, which must be considered when assessing their expected performance on observational data.}

{The 2D Gaussian fitting method has the weakest dependence on Monte Carlo simulations, as it does not require simulation-based training. Instead, it directly corrects the geometrical bias caused by image truncation at the camera edge, making it particularly useful for events with large leakage. }

{The LightGBM-based uncertainty weighting has a stronger dependence on Monte Carlo simulations, since the reconstruction quality of each telescope image is estimated using a model trained on simulated events. However, the predicted uncertainties are used mainly as relative weights between telescopes, so an accurate absolute calibration is not essential. This may make the method more robust against moderate differences between simulations and observations. }

{The  likelihood-based reconstruction has the   strongest dependence on the accuracy of the simulations. In this approach, the observed pixel-wise images are compared directly with simulation-derived probability models. Consequently, mismodelling of the atmosphere, optical point spread function, detector calibration, electronic response, or night-sky background (NSB) can propagate directly into the reconstructed shower parameters. A detailed evaluation of these systematic effects will therefore be necessary before the performance obtained from Monte Carlo simulations can be considered representative of real LACT observations.}

We summarize the performance improvements and characteristics of the different methods in Table~\ref{tab:method_comparison}. As an initial robustness test, we investigated the influence of increased night-sky background (NSB). The same LightGBM and likelihood models developed under nominal dark-sky conditions were applied, without retraining, to simulations with NSB levels increased to 200\% and 300\% of the nominal value. The same image-cleaning thresholds were also retained. For both methods, the improvement relative to the standard reconstruction remains largely stable across the tested NSB levels, indicating that their performance is not strongly degraded by this particular change in observing conditions. The corresponding performance improvements are also summarized in Table~\ref{tab:method_comparison}.

\subsection{Broader implications and future developments}

Beyond the specific performance gains obtained for LACT, the methods investigated in this work may also have broader implications for stereoscopic IACT reconstruction. The 2D Gaussian fitting method is particularly relevant for observations in which image truncation becomes significant. Since it does not require simulation-based training, it can be readily adapted to other IACTs with relatively small fields of view, such as VERITAS, MAGIC, and the LSTs.

The uncertainty-weighting approach may be especially useful for other next-generation IACT arrays with large telescope multiplicities at high energies, such as CTA and ASTRI. In such events, the reconstruction quality can vary substantially among individual telescope images, and appropriately weighting their contributions provides a natural way to exploit the additional stereoscopic information.

For the pixel-wise likelihood reconstruction, two main directions could be explored for further improvement. First, incorporating uncleaned, noisy pixels into the likelihood model could recover sub-threshold information, particularly in regions where a signal is physically expected but falls below conventional image-cleaning thresholds. Second, while the current formulation assumes statistical independence between pixels, neighboring pixels naturally contain correlated spatial information. Capturing these correlations could provide a more complete description of the shower image and potentially lead to further improvements in reconstruction performance. Future developments may address this complexity using more flexible representation-learning techniques, such as Masked Autoencoders (MAE; \cite{he2022masked}), to implicitly learn the joint spatial distribution of pixel signals.
\subsection{Event-type analysis}

A further extension of the methods developed in this work is their ability to estimate reconstruction quality on an event-by-event basis. In particular, the likelihood-based framework can naturally extend this information to other reconstructed quantities, such as energy. Inspired by the event-type analysis used by Fermi-LAT \cite{abdo2010fermi}, events could be divided into different quality classes according to their estimated uncertainties. This would allow different subsets of the data to be optimized for different scientific goals, for example selecting events with the best angular resolution for morphological studies, or events with the best energy resolution for searches for spectral features \cite{hassan2021performance}. Such an approach could make more efficient use of the available events and further improve the scientific performance of LACT.




\section{Summary}

In this work, we presented a systematic study of stereoscopic direction
reconstruction for the LACT array, with particular emphasis on the degradation
of angular resolution at high energies and large source offsets. By comparing
the standard LACT camera with a circular-camera configuration, we identified
image truncation at the camera edge as a major source of reconstruction bias.
For strongly truncated images, the standard Hillas parametrization can produce
a biased estimate of the image major axis, particularly for high-energy events
with large impact parameters.

To mitigate this effect, we introduced a hybrid reconstruction in which a
2D Gaussian fit is used for images with large leakage. This approach
substantially improves the reconstruction of truncated images while retaining
the standard Hillas parametrization for well-contained images. At
$100~\mathrm{TeV}$, the hybrid reconstruction improves the angular resolution
by approximately $0.03^\circ$--$0.04^\circ$ across most offset ranges. In the
outer $3^\circ$--$4^\circ$ offset bin, the improvement reaches approximately
25--30\%.

We further investigated uncertainty-weighted stereoscopic reconstruction, in
which the estimated reconstruction quality of individual telescope images is
used to weight their contributions. Both the \textit{HillasWeightedSum} and
\textit{HillasWeightedDisp} methods improve the angular resolution in the
high-energy and large-offset regime, with improvements of up to
$0.02^\circ$--$0.03^\circ$ relative to the hybrid Gaussian reconstruction.
The \textit{HillasWeightedDisp} method additionally provides improved
performance at low energies, where individual telescope images generally
contain fewer signal pixels and have poorer reconstruction quality.

Finally, we explored a pixel-wise likelihood reconstruction based on neural
ratio estimation. By directly exploiting the pixel-level image information,
this method achieves an angular resolution better than $0.09^\circ$ at
$100~\mathrm{TeV}$ across the full field of view in the simulations considered
here. Among the approaches investigated in this work, it provides the largest
potential improvement, although its performance depends more strongly on the
accuracy of the Monte Carlo description, as discussed above.

Overall, this study demonstrates that improved reconstruction methods can substantially enhance the angular resolution of LACT, particularly for high-energy and off-axis observations.

\appendix
\section{Features Used in LightGBM}\label{app:parameter}

This appendix provides detailed descriptions of the features utilized in the LightGBM model for event reconstruction and classification.

\subsection{Hillas Parameters}
\begin{itemize}
    \item {hillas\_width}: The width (minor axis) of the Hillas ellipse.
    \item {hillas\_length}: The length (major axis) of the Hillas ellipse.
    \item {hillas\_intensity}: The total number of photoelectrons (p.e.) detected within the cleaned image.
    \item {hillas\_r}: The distance between the center of the camera and the Center of Gravity (CoG) of the ellipse.
\end{itemize}

\subsection{Leakage Parameters}
\begin{itemize}
    \item {leakage\_pixels\_width\_1}: The fraction of pixels located in the outermost ring of the camera.
    \item {leakage\_pixels\_width\_2}: The fraction of pixels located within the two outermost rings of the camera.
    \item {leakage\_intensity\_width\_1}: The fraction of total intensity (p.e.) contained in the outermost ring of pixels.
    \item {leakage\_intensity\_width\_2}: The fraction of total intensity (p.e.) contained within the two outermost rings of pixels.
\end{itemize}

\subsection{Concentration and Intensity Parameters}
\begin{itemize}
    \item {concentration\_pixels}: The ratio of the intensity of the brightest pixels to the total intensity.
    \item {concentration\_cog}: The fraction of intensity contained within a one-pixel diameter of the CoG.
    \item {concentration\_core}: The fraction of intensity located inside the Hillas ellipse, defined by the region:
    \[ \frac{x'^{2}}{\sigma_L^2} + \frac{y'^{2}}{\sigma_W^2} < 1 \]
    \item {intensity\_mean}: The mean number of photoelectrons per image pixel.
    \item {intensity\_std}: The standard deviation of photoelectrons across image pixels.
    \item {intensity\_max}: The intensity of the brightest pixel in the image.
\end{itemize}

\subsection{Morphology and Fitting}
\begin{itemize}
    \item {morphology\_num\_pixels}: The total number of pixels contained within the cleaned image.
    \item {rec\_impact\_parameter}: The reconstructed impact parameter.
    \item {use\_gaussian\_fit}: A boolean flag indicating whether  Gaussian fit was used.
    \item {gaussian\_fit\_size}: The total intensity as determined by the Gaussian fitting procedure.
\end{itemize}
\section{Analytical Solution for the \textit{HillasWeightedSum} Method}

In the \textit{HillasWeightedSum} method, the objective is to determine the optimal source position $(x, y)$ by minimizing the sum of the squared weighted perpendicular distances from this point to the major axes of the participating telescopes. The cost function $J(x, y)$ is defined as:

\begin{equation}
    J(x,y) = \sum_i w_i d_i^2,
\end{equation}
where $w_i = 1/\sigma_{\mathrm{miss}, i}^2$ represents the weight derived from the uncertainty, and $d_{i}$ denotes the perpendicular distance between the point $(x, y)$ and the major axis of the $i$-th telescope.

The major axis of the telescope image is defined by the line equation passing through the center of gravity $(x_{\mathrm{cog}, i}, y_{\mathrm{cog}, i})$ with an orientation angle $\alpha_i$:
\begin{equation}
    -\sin\alpha_i (x - x_{\mathrm{cog}, i}) + \cos\alpha_i (y - y_{\mathrm{cog}, i}) = 0.
\end{equation}

This can be rewritten in the general linear form $d_i = a_i x + b_i y + c_i$, where the coefficients are:
\begin{equation}
    \begin{aligned}
        a_i &= -\sin\alpha_i, \\
        b_i &= \cos\alpha_i, \\
        c_i &= x_{\mathrm{cog}, i} \sin\alpha_i - y_{\mathrm{cog}, i} \cos\alpha_i.
    \end{aligned}
\end{equation}

Thus, the minimization problem becomes:
\begin{equation}
    J(x, y) = \sum_i w_i (a_i x + b_i y + c_i)^2.
\end{equation}

To find the minimum, we set the partial derivatives of $J(x, y)$ with respect to $x$ and $y$ to zero:
\begin{equation}
    \begin{aligned}
        \frac{\partial J}{\partial x} &= 2 \sum_i w_i a_i (a_i x + b_i y + c_i) = 0, \\
        \frac{\partial J}{\partial y} &= 2 \sum_i w_i b_i (a_i x + b_i y + c_i) = 0.
    \end{aligned}
\end{equation}

Rearranging these linear equations allows us to express the system in matrix form $ \mathbf{M} \mathbf{r} = \mathbf{V} $:

\begin{equation}
    \begin{bmatrix}
        \sum w_i a_i^2 & \sum w_i a_i b_i \\
        \sum w_i a_i b_i & \sum w_i b_i^2
    \end{bmatrix}
    \begin{bmatrix}
        x \\
        y
    \end{bmatrix}
    = -
    \begin{bmatrix}
        \sum w_i c_i a_i \\
        \sum w_i c_i b_i
    \end{bmatrix}.
\end{equation}

Solving this linear system provides the optimal analytical solution for the source position $(x, y)$.
\label{app1}

\section{Analytical Solution for the \textit{HillasWeightedDisp} Method}

For a detailed discussion on the \textit{HillasWeightedDisp} method, one can refer to \cite{stamatescu2010new}. Here, we present the derivation for combining error ellipses from a probabilistic perspective.

Assuming each telescope $i$ provides an independent measurement of the source position, the estimated coordinate vector $\boldsymbol{\mu}_i$ is given by the center of gravity $(x_{\mathrm{cog},i}, y_{\mathrm{cog},i})$ projected along the major axis by the \textit{disp} parameter $\delta_i$:
\begin{equation}
    \boldsymbol{\mu}_i = 
    \begin{pmatrix} 
        \mu_{x,i} \\ 
        \mu_{y,i} 
    \end{pmatrix} 
    = 
    \begin{pmatrix} 
        x_{\mathrm{cog},i} + disp \cdot \cos\alpha_i \\ 
        y_{\mathrm{cog},i} + disp \cdot \sin\alpha_i 
    \end{pmatrix}.
\end{equation}

The uncertainty of this measurement is represented by a covariance matrix $\Sigma_i$. If we define the intrinsic covariance matrix in the telescope frame  as $\mathbf{C}_i = \operatorname{diag}(\sigma_{disp,i}^2, \\\sigma_{miss,i}^2)$, the covariance in the camera coordinate system is obtained by applying a rotation matrix $\mathbf{R}(\alpha_i)$:
\begin{equation}
    \Sigma_i = \mathbf{R}(\alpha_i) \mathbf{C}_i \mathbf{R}(\alpha_i)^\top,
\end{equation}
where the rotation matrix is:
\begin{equation}
    \mathbf{R}(\alpha_i) = 
    \begin{pmatrix} 
        \cos\alpha_i & -\sin\alpha_i \\ 
        \sin\alpha_i & \cos\alpha_i 
    \end{pmatrix}.
\end{equation}

The probability density function for the source position $\mathbf{r} = (x, y)^\top$, given the measurement from telescope $i$, follows a 2D normal distribution:
\begin{equation}
    P_i(\mathbf{r} \mid \boldsymbol{\mu}_i) = \frac{1}{2 \pi \sqrt{\operatorname{det}(\Sigma_i)}} \exp \left(-\frac{1}{2}(\mathbf{r}-\boldsymbol{\mu}_i)^{\top} \Sigma_i^{-1} (\mathbf{r}-\boldsymbol{\mu}_i)\right).
\end{equation}

Assuming the observations from all $N$ telescopes are statistically independent, the joint likelihood function $L(\mathbf{r})$ is the product of the individual probabilities:
\begin{equation}
    L(\mathbf{r}) = \prod_{i=1}^{N} P_i(\mathbf{r} \mid \boldsymbol{\mu}_i).
\end{equation}

To find the optimal source position, we maximize the likelihood $L(\mathbf{r})$, which is equivalent to minimizing the negative log-likelihood function (often denoted as the $\chi^2$ or loss function $J$):
\begin{equation}
    J(\mathbf{r}) = -2 \ln L(\mathbf{r}) = \sum_{i=1}^{N} (\mathbf{r}-\boldsymbol{\mu}_i)^{\top} \mathbf{W}_i (\mathbf{r}-\boldsymbol{\mu}_i) + \text{const},
\end{equation}
where $\mathbf{W}_i = \Sigma_i^{-1}$ represents the weight matrix for telescope $i$.

To minimize $J(\mathbf{r})$, we calculate the gradient with respect to $\mathbf{r}$ and set it to zero:
\begin{equation}
    \nabla_{\mathbf{r}} J = 2 \sum_{i=1}^{N} \mathbf{W}_i (\mathbf{r} - \boldsymbol{\mu}_i) = 0.
\end{equation}

This leads to the linear matrix equation:
\begin{equation}
    \left( \sum_{i=1}^{N} \mathbf{W}_i \right) \mathbf{r} = \sum_{i=1}^{N} \mathbf{W}_i \boldsymbol{\mu}_i.
\end{equation}

Finally, the optimal source position $\mathbf{r}_{\mathrm{est}}$ is obtained analytically by inverting the sum of the weight matrices:
\begin{equation}
    \mathbf{r}_{\mathrm{est}} = \left( \sum_{i=1}^{N} \mathbf{W}_i \right)^{-1} \left( \sum_{i=1}^{N} \mathbf{W}_i \boldsymbol{\mu}_i \right).
\end{equation}
\label{app2}

\section*{Acknowledgments}

Rui-zhi Yang is supported by the NSFC under grants 12588101 and 12393854, and by the Natural Science Foundation of Sichuan Province under grant 2025ZNSFSC0065. Rui-zhi Yang gratefully acknowledges the support of the Cyrus Chun Ying Tang Foundation and the Studio of Academician Zhao Zhengguo, Deep Space Exploration Laboratory. Jiali Liu is supported by the NSFC (grant no. 12573107). Shoushan Zhang is supported by the Sichuan Province Science Foundation for Distinguished Young Scholars under grant no. 2022JDJQ0043, and the National Natural Science Foundation of China under grant no. 12261141691.

\section*{Data Availability Statement}

The Monte Carlo simulation data used in this study are internal to the LACT Collaboration and are not publicly available due to the collaboration's data policy.
%
\bibliographystyle{unsrt}
\bibliography{ref}

\end{document}